# Vacancy-induced pseudo-gap formation in antiferromagnetic Cr$_{0.86}$ZnSb

Michael Parzer[1*], Fabian Garmroudi[1], Herwig Michor[1], Xinlin Yan[1], Ernst Bauer[1], Gerda Rogl[2], Jiri Bursik[3], Stephen Cottrell[4], Raimund Podloucky[2], Peter Rogl[2]

[1] Institute of Solid State Physics, TU Wien, Wiedner Hauptstr. 8-10, A-1040 Wien, Austria
[2] Institute of Materials Chemistry, Universität Wien, Währingerstr. 42, A-1090 Wien, Austria
[3] Institute of Physics of Materials, Czech Academy of Sciences, Žižkova 22, 61662 Brno, Czech Republic
[4] ISIS Facility, Rutherford Appleton Laboratory, Chilton, Didcot Oxon OX11 0QX, United Kingdom

**Abstract**

Structural defects are important for both solid-state chemistry and physics, as they can have a significant impact on chemical stability and physical properties. Here, we identify a vacancy-induced pseudo-gap formation in antiferromagnetic Cr$_{0.86}$ZnSb. Cr$_{1-x}$ZnSb alloys were studied combining efforts of density functional theory (DFT) calculations and experimental methods to elucidate the effect of vacancies. Detailed analyses (X-ray powder and single crystal diffraction, transmission and secondary scanning electron microscopy) of Cr$_{1-x}$ZnSb, $0<x<0.20$, prompts Cr$_{0.86}$ZnSb as the only stable compound, crystallizing with the MnAlGe-type structure. From DFT calculations, an antiferromagnetic spin configuration of Cr local magnetic moments was found to be favorable for both, the perfectly stoichiometric compound CrZnSb, as well as for Cr$_{0.875}$ZnSb. Magnetic order is observed experimentally for Cr$_{0.86}$ZnSb by temperature- and field-dependent magnetization measurements, revealing a magnetic phase transition near 220 K, which is corroborated by zero-field muon spin relaxation studies. Thermoelectric transport properties exhibit distinct maxima in the temperature-dependent Seebeck coefficient and electrical resistivity at around 190 K. Analyzing the measured data on the basis of a triple parabolic band model and DFT simulations, their characteristic features are traced back to a pseudo-gap in the electronic structure arising from a particular vacancy arrangement. These findings offer valuable insights into the role of vacancies in defected materials, contributing to the broader understanding of structural defects and their impact on the electronic structure.

1. **Introduction**

Structural defects, particularly vacancies, are known to profoundly influence the macroscopic physical properties of materials, a phenomenon that has been extensively studied in half-Heusler compounds [1-3]. In the field of thermoelectrics, vacancies have been reported to enhance transport properties in specific cases [4-6]. Therefore, a profound understanding of the role of structural defects in these materials is essential for the effective tuning of their electronic properties [7-9]. Moreover, vacancies in half-Heusler structures have been shown to induce long-range magnetic order, similar to the effects observed in oxides [2,10,11].

Half-Heusler alloys are a relevant group of thermoelectric (TE) materials with a promising figure of merit $ZT$ and they have drawn attention due to the discovery of a band gap in {Ti,Zr,Hf}NiSn compounds and impurity states within the gap [12-14]. Record values of the thermoelectric figures of merit were revealed for Sb-doped $Ti_{0.5}Zr_{0.25}Hf_{0.25}NiSn$ which reaches $ZT_{max} \sim 1.5$ [15,16]. Interestingly, the 3d-metal series forms $T$ZnSb compounds with different structures depending on the transition metal $T$. For more electropositive elements $T$ = Ti, V, Cr, Mn, Fe, the compounds typically adopt the MnAlGe-type structure. For transition metals with nearly filled 3d shells, including Fe at higher temperatures, the compounds crystallize in the half-Heusler type structure (FeZnSb [17,18], NiZnSb [19]) or form a 2×2×2 superstructure of the half-Heusler unit cell ($T$ = Fe, Co) $Ru_9Zn_7Sb_8$-type [20]). Quite recently Benson et al. [21] reported on disordered $V_{1-x}ZnSb_{1-y}$ ($x \sim 0.1$, $y \sim 0.05$), for which itinerant charge carriers were said to coexist with a small, localized magnetic moment of around 0.25 μB from the ordered part of the vanadium lattice (MnAlGe-type). In a follow-up paper, structural and thermodynamic properties, as well as resistivity were reported for $X$ZnSb phases with $X$ = Cr, Mn, Fe, Ni for which sizable amounts of vacancies on transition-metal sites were revealed [22]. Given the intriguing physical properties of $Cr_{1-x}ZnSb$, including a resistivity maximum near room temperature and unresolved magnetic ordering [22], we aim to scrutinize the structure-property relationship in this system, with a particular focus on how the Cr vacancies influence these properties. To analyze transport and magnetic properties using density-functional theory (DFT) calculations, experimental results from phase-pure samples must be compared with respect to appropriate supercell calculations.

Accordingly, we focus our investigations on several tasks: (i) to synthesize single phase $Cr_{1-x}ZnSb$, (ii) to elucidate the crystal structure and defect sites, (iii) to reveal physical properties, including thermal stability, specific heat, electrical conductivity, Seebeck coefficient and magnetic behavior, and (iv) by elaborate DFT calculations to provide detailed information about the effect of the vacancies on structural stability, magnetic ordering and magnetic moments, on electronic structure and bonding, and on electronic transport properties.

## 2. Methods

### 2.1. Experimental techniques

Due to the rather high vapour pressures of Zn and Sb at elevated temperatures, arc melting was avoided and several samples Cr$_{1-x}$ZnSb ($x$ = 0, 0.05, 0.1, 0.14, 0.15, 0.2) were prepared via conventional powder metallurgical sinter techniques from powders of 99.9 mass% Cr-powder (<200 mesh; Alfa Ventron), filings of a 99.8 mass% pure Zn-rod (Alfa Ventron) and freshly powdered 99.99 Sb-rod (in a WC-Co mortar) from Alfa Ventron, Germany. The powder blends were compacted without lubricants in a steel die and sealed in a quartz tube. Low thermal stability demanded individual reaction temperatures and therefore, proper annealing temperatures were extracted from DTA: Cr$_{1-x}$ZnSb proved to be stable only below <660°C; thus samples were annealed at 550°C. After each second day (for 5 times) the sinter cake was powdered (< 60 µm), re-compacted and sealed for annealing again. With this procedure rather single-phase products (3 to 5 grams) could be obtained. As in all cases the sinter cake appeared porous, densification was achieved (relative density about 90 %) for cylindrical plates ($\varnothing$ = 10 mm, height ~1 mm) via high pressure torsion (HPT) at room temperature in steel dies at 4 GPa and up to 5 revolutions. The measured density, d$_A$, was derived from Archimedes' principle on the hot-pressed cylinders in distilled water. The X-ray-density d$_X$ = MZ/VL was calculated from Z as the number of formula units within the unit cell, M as the molar mass in g/mol, V as the volume and L as Loschmidt's number, the relative density, d$_{rel}$, was calculated from d$_{rel}$ = (d$_A$/d$_X$)×100.

Polycrystalline powders were analyzed by X-ray powder diffraction with Ge-monochromated Cu-$K\alpha_1$ radiation ($\lambda$ = 0.154056 nm) employing a Guinier-Huber image plate recording system. Rietveld refinements were performed using the program FULLPROF [23], whilst precise lattice parameters were obtained from least squares fits with the program STRUKTUR [24]; Ge served as an internal standard (a$_{Ge}$ = 0.5657906 nm).

The analysis of SEM microstructures and X-ray patterns of Cr$_{1-x}$ZnSb revealed that only for $x$ = 0.14, a phase pure compound was observed. For the remaining values of x, impurity phases with different concentrations were obtained. Thus, the following part of the manuscript focuses on Cr$_{0.86}$ZnSb, and only data and analyses of this sample are presented. A rather spherical single crystal fragment in the range of 30 to 60 µm was isolated from the alloy Cr$_{0.86}$ZnSb, which was additionally heated to 650°C and slowly cooled to 500°C. X-ray intensity data were collected at room temperature on a four-circle APEX II diffractometer [25] equipped with a CCD area detector and an Incoatec Microfocus Source IµS (30 W, multilayer mirror, Mo-K$_\alpha$; $\lambda = 0.071069$ nm; detector distance of 40 mm; full sphere; $2° < 2\theta < 70°$).

All samples were ground on SiC papers and polished with Al$_2$O$_3$ powders (down to 0.3 µm) via standard procedures to be examined by light optical metallography (LOM) and scanning electron microscopy (SEM). The microstructure and chemical composition were analyzed by a Jeol JSM-6460 scanning electron microscope equipped with an energy dispersive X-ray (EDX) detector operated at 20 kV and by a Tescan LYRA 3XMH FEG/SEM scanning electron microscope equipped with an X-Max80 Oxford Instruments energy dispersive X-ray (EDX) detector operated at 20 kV. Quantitative evaluation of compositions was performed with the INCA - software [26].

A thin lamella (lateral dimensions of about 9×6 µm$^2$) was prepared from the polished sample Cr$_{0.85}$ZnSb using a focused ion beam (FIB) technique in a TESCAN LYRA 3 XMU FEG/SEM×FIB scanning electron microscope. A ThermoFisher Scientific Thalos F200i transmission electron microscope (TEM) with FEG operated at 200 kV was employed in diffraction mode, to support information about the crystal structure.

The electrical resistivity in the range from 4.2 to 300 K, was measured on cuboids (8×2×1 mm³) in a conventional $^4$He cryostat via a four-probe technique, employing an a.c. Lake Shore Resistance Bridge 370 AC. The Seebeck coefficient at low temperatures was obtained using the see-saw heating method, as described in Ref. [27]. Above room temperature, the electrical resistivity and the Seebeck coefficient were measured simultaneously with a ZEM-3 equipment (ULVAC-Riko, Japan). The error for both, resistivity measurements as well as for the Seebeck coefficient, is about 5 %. For specific heat measurements, a Quantum Design PPMS was employed for temperatures from 2 to 300 K. Apiezon-N grease served to ensure a good thermal contact between sample and sample platform. Temperature- and field-dependent magnetization data were collected at temperatures ranging from 2 to 350 K using a Quantum Design PPMS with the vibrating sample magnetometer (VSM) option.

Muon spin relaxation (µSR) experiments were performed with the EMU spectrometer at the ISIS neutron and muon spallation source at the Rutherford Appleton Laboratory, UK [28]. Muons are implanted and thermalized within few ps in the Cr$_{0.86}$ZnSb powder sample and a minor fraction in the Ag sample holder and decay with a half-life, $\tau_\mu = 2.2$ µs into positrons. The latter are emitted with an asymmetric probability with respect to the muon spin orientation. The muon spin polarisation, thus, corresponds to the measured asymmetry between positron counts in forward and backward detectors (in so-called longitudinal detector geometry). Based on a statistic of typically $2 \cdot 10^7$ muon decays, the asymmetry is determined as a function of time. Zero-field µSR measurements were taken in temperature steps of 10 K ranging from 20 K to 240 K and at 300 K. At the latter temperature, additional 20 G transverse field data were collected to calibrate the relative efficiency of the forward and backward detectors. Accordingly, an initial asymmetry of approximately 27 % refers to a perfect spin polarization of the implanted muons parallel to the beam forward direction. We used the WIMDA software [29] for analysing the µSR data.

### 2.2 Computational aspects

For the density functional theory (DFT) calculations, the Vienna *ab initio* simulation package (VASP) [30,31] with the projector augmented wave potential construction (PAW) [32,33] was applied. The Zn pseudopotential was constructed by treating the $3d^{10}$ and $4s^2$ states as valence states, whereas for Sb the pseudopotential included the $5s^2$ and $5p^3$ states as valence states. For the Cr pseudopotential construction, the atomic orbital configuration $3s^2 3p^6 3d^5 4s^1$ was used. A plane wave basis cutoff of 600 eV was chosen, ensuring highly converged results with respect to basic functions. Dense *k* point grids were constructed by setting the VASP parameter KSPACING = 0.06 in all calculations. For example, for the perfect $P4/nmm$ structure of $Cr_2Zn_2Sb_2$ it resulted in a grid of $25 \times 25 \times 17$ *k* points. For the derivation of local (or site projected) properties such as $\ell$-projected densities of states, magnetic moments, and atom-like charges in atomic spheres of radii 0.116 nm for Cr, 0.124 nm for Zn, and 0.158 nm for Sb were circumscribed each corresponding atomic position. All crystal structure parameters were fully relaxed by minimising total energies and atomic forces. Numerous DFT spin-polarised calculations were performed for studying structural stabilities and magnetic structures. All DFT results were obtained employing the standard GGA-PBE ansatz [34] for exchange and correlation without any corrections for additional enhanced exchange-correlation interactions such as the DFT+U approach of Dudarev et al. [35]. This approach overestimates the equilibrium volumes and consequently leads to less agreeable results (see chapter 3.3). Charge transfer, atomic volumes and atom-like charges were computed by analysing the charge density in terms of the quantum theory of atoms in molecules according to Bader et al. [36-41]. Applying an adapted version of the BoltzTrap package [42], thermoelectric properties such as the Seebeck coefficient and electrical resistivity were computed within Boltzmann's transport theory assuming constant relaxation time.

## 3. Synthesis and characterization

### 3.1. Structure and structure stability of Cr$_{0.86}$ZnSb.

The X-ray intensity patterns of a single crystal specimen, extracted from the rim of a sample of Cr$_{0.86}$ZnSb, were fully indexed and were unambiguously consistent with the space group *P*4/*nmm* (No 129, origin at centre) and lattice parameters: a = 0.41704(1) nm and c = 0.61855(2) nm; c/a = 1.483. Structure solution by direct methods indicates isotypism with the structure type of MnAlGe (anti-PbFCl-type) as defined for MnZnSb (for details see Pearson's Crystal Data compilation [43]). The refinement, however, documents a significant amount of random defects on the Cr-site (0.86(1) Cr in 2a (¾,¼,0)), a rather small, almost insignificant, amount of defects in the Zn-site 2c (¼, ¼, z = 0.27985(10); occ = 0.98(1)) and full occupancy on the Sb-site in 2c (¼, ¼, z = 0.71853(5)). Final refinements with anisotropic atom displacement parameters (ADPs) converged to $R_F = 0.0124$ with residual electron densities smaller than ± 1.37 e$^-$ × 10$^{-3}$ nm$^3$. The crystallographic data are summarized in Table 1a,b, including interatomic distances, which are consistent with the sum of CN12 metal atom radii (CN is the coordination number; $R_{Cr}$ = 0.1360 nm, $R_{Zn}$ = 0.1394 and $R_{Sb}$ = 0.1590 nm [44]). A quantitative Rietveld refinement of the X-ray powder intensities of Cr$_{0.86}$ZnSb ($R_F$ = 0.055; see Fig. 1) clearly confirms the structure as derived from the single crystal study, as well as the Cr-defects. The structure model of Cr$_{0.86}$ZnSb (≡ Cr$_{30.1}$Zn$_{34.5}$Sb$_{35.2}$ at.%) is satisfactorily refined and in excellent agreement with the composition of Cr$_{30.9}$Zn$_{33.8}$Sb$_{35.3}$ at.%, as derived from electron probe microanalysis (EPMA) (see insert in Fig. 1). It should be emphasized that the least squares structure refinement did neither accept extra Sb at the Cr-site nor at the Zn-site.

The structure of Cr$_{0.86}$ZnSb, as shown in Fig. 1, can be viewed as a sequence of densely packed but defect Cr-layers in a,b planes (with direct Cr-Cr distances d$_{Cr-Cr}$ = 0.29489 nm) stacked along the c-axis with two Zn/Sb layers in between. While Cr-Zn, Cr-Sb and particularly Zn-Sb distances are consistent with the sum of radii, the structural arrangement also induces a long distance for the Cr-Cr interaction along the c-direction (0.61855(2) nm). This geometry may severely affect the magnetic properties (see below). Coordination figures are simple tricapped triangular prisms for Zn and Sb (coordination number CN = 9) and a cuboctahedron for Cr (CN = 12). No homogeneity region in direction of a full Cr-occupancy was detected.

Table 1a: X-ray single crystal and powder data for Cr$_{0.86}$ZnSb; MnAlGe-type, space group *P4/nmm*, Nr. 129 (origin at centre); The crystal structure was solved with programs SHELXS-97, SHELXL-97-2 within the program OSCAIL [45]. Finally, the crystal structure was standardized with the program Structure Tidy [46].

| Parameter/compound | Cr$_{1-x}$ZnSb (Rietveld) | Cr$_{1-x}$ZnSb (single crystal) |
|---|---|---|
| Composition from EPMA in at.% | Cr30.9Zn33.8Sb35.3 | Cr30.9Zn33.8Sb35.3 |
| Formula from refinement in at.% | Cr30.1Zn34.9Sb35 | Cr30.1Zn34.5Sb35.2 |
| Formula from refinement | Cr$_{0.86}$ZnSb | Cr$_{0.86}$Zn$_{0.98}$Sb |
| Radiation | Cu K$\alpha$1 (Guinier) | Mo K$\alpha$ (APEX II) |
| $2\theta$ range (degrees) | $18 \leq 2\theta \leq 100$ | $2 \leq 2\theta \leq 72.7$ |
| Linear absorption coeff. in mm$^{-1}$ | - | 27.16 |
| Density in Mg/m$^3$ | 7.34 | 7.13 |
| $a = b$, $c$; all in nm | 0.41797(1); 0.61944(1) | 0.41704(1); 0.61855(2) |
| Reflections in refinement (total) | 51 | 178 Fo > 4$\sigma$ (Fo) (5710) |
| Number of variables | 24 | 12 |
| $R_F = \Sigma\|F_o-F_c\|/\Sigma F_o$ | 0.0549 | 210 frames (5 sets) |
| $R_I = \Sigma\|I_o-I_c\|/\Sigma I_o$ | 0.0760 | Redundancy > 8 |
| $R_{wP} = [\Sigma w_i\|y_{oi}-y_{ci}\|^2/\Sigma w_i\|y_{oi}\|^2]^{1/2}$ | 0.0412 | $R_F^2 = \Sigma\|F_0^2-F_c^2\|/\Sigma F_0^2 = 0.0124$ |
| $R_P = \Sigma\|y_{oi}-y_{ci}\|/\Sigma\|y_{oi}\|$ | 0.0618 | $wR2 = 0.0288$ |
| $R_e = [(N-P+C)/(\Sigma w_i y^2_{oi})]^{1/2}$ | 0.0182 | $R_{int} = 0.026$ |
| $\chi^2 = (R_{wP}/R_e)^2$ | 11.5 | GOF = 1.154 |
| Extinction (Zachariasen) | - | 0.044(2) |
| Residual density e$^-$/nm$^3 \times 10^3$; max; min. | - | 1.37 (0.065 nm from Zn), -0.93 |
| **Atom positions** | | |
| Cr: 2a (¾,¼,0); Occ. | 0.86(1) Cr | 0.863(4) Cr |
| $U_{11}=U_{22}$; $U_{33}$; $U_{23}=U_{13}=U_{12}=0$ | Biso=0.47 | 0.0131(3); 0.0067(3) |
| Zn: 2c (¼, ¼, z); Occ. | 1.0(-) Zn | 0.983(4) Zn |
| z | 0.2814(4) | 0.27985(10) |
| $U_{11=}U_{22}$; $U_{33}$; $U_{23}=U_{13}=U_{12}=0$ | Biso=0.59 | 0.0163(2); 0.0176(3) |
| Sb: 2c (¼, ¼, z); Occ. | 1.00(-) Sb | 1.00(-) Sb |
| z | 0.7175(2) | 0.71853(5) |
| $U_{11=}U_{22}$; $U_{33}$; $U_{23}=U_{13}=U_{12}=0$ | Biso=0.32 | 0.0104(1); 0.0101(1) |

Table 1b: Interatomic distances (< 0.4 nm) for single crystal data of Cr$_{0.86}$ZnSb.

| Cr | - | 4 Zn | 0.27101 |
|---|---|---|---|
|  | - | 4 Sb | 0.27164 |
|  | - | 4 Cr | 0.29489 |
| Zn | - | 4 Cr | 0.27101 |
|  | - | 1 Sb | 0.27135 |
|  | - | 4 Sb | 0.29489 |
| Sb | - | 1 Zn | 0.27135 |
|  | - | 4 Cr | 0.27165 |
|  | - | 4 Zn | 0.29489 |

Selected area electron diffraction (SAED) patterns were collected by TEM from thin lamellae of the Cr$_{0.86}$ZnSb sample (see Fig. 1, lower panel). All diffraction patterns were fully indexed using the parameters of the tetragonal lattice obtained from the single crystal refinement (Table 1 and text above). No further superstructure spots were detected - thus no ordering exists among the randomly distributed Cr-atoms and vacancies. The lower panel in Fig. 1 shows several examples of low index zone axis diffraction patterns recorded at various sample tilts together with simulated SAED patterns using the software JEMS [47,48].

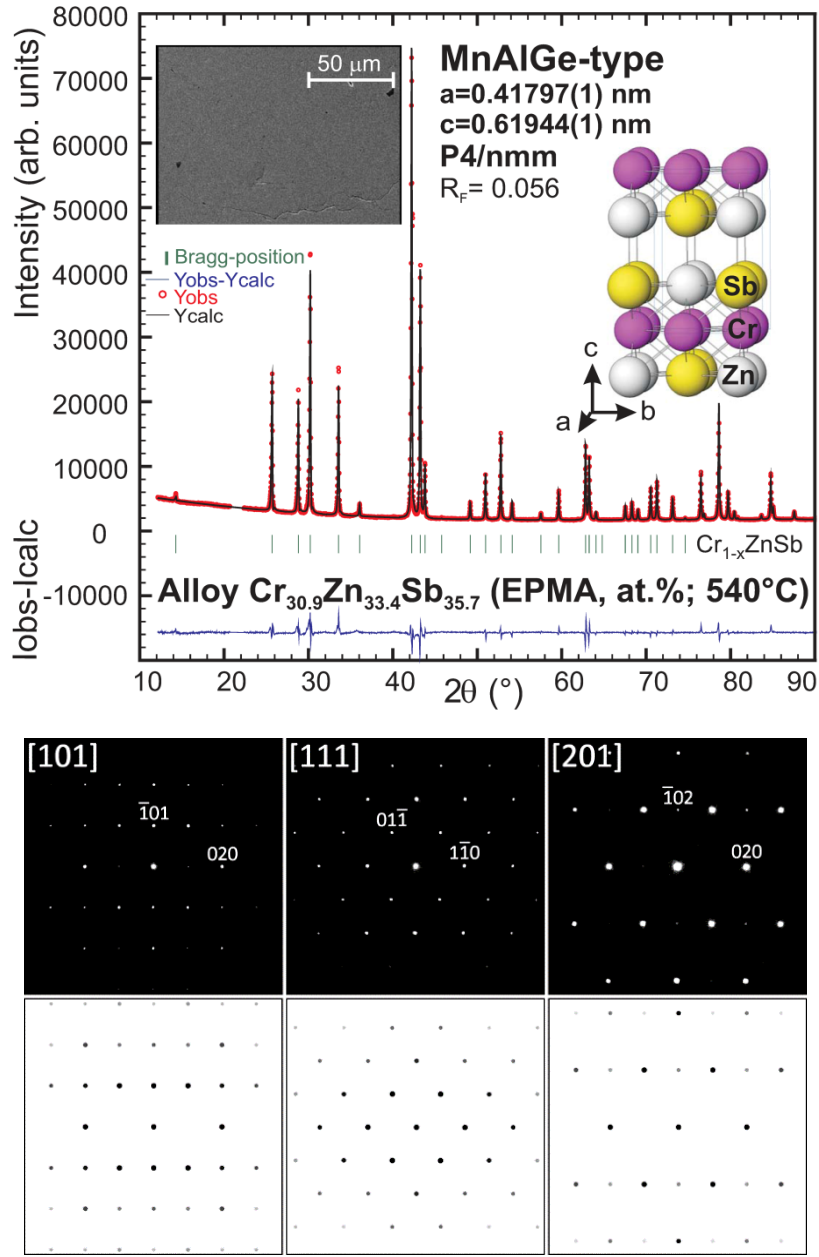

Figure 1: Upper panel: Rietveld refinement for Cr$_{1-x}$ZnSb; x = 0.14 (Inset left: EPMA backscatter micrograph; Inset right: structure of Cr$_{0.86}$ZnSb in three-dimensional view: Cr-atoms violet, Zn-atoms light grey, Sb-atoms yellow). Lower panel: Selected area electron diffraction (SAED) patterns of Cr$_{0.86}$ZnSb at various sample tilts together with the results of kinematic simulation.

### 3.2. DFT structures for Cr$_2$Zn$_2$Sb$_2$ (C-AF) and Cr$_4$Zn$_4$Sb$_4$ (G-AF)

In the following chapter, two types of DFT-derived anti-ferromagnetic ground states (C-AF and G-AF) are discussed. Both orderings feature a checker-board pattern in the ($\vec{a}, \vec{b}$) plane, depicted in Fig. 2b [11]. The difference between the two structures is the ordering in $\vec{c}$ direction. For C-AF, the Cr moments are ordered parallel, while for G-AF ordering, an alternating up/down magnetic ordering occurs along $\vec{c}$, doubling the unit cell in $\vec{c}$ direction (see Fig. 2a, b). For the C-AF structure of Cr$_2$Zn$_2$Sb$_2$, antiferromagnetic ordering occurs in the ($\vec{a}, \vec{b}$) plane (Fig. 2b).

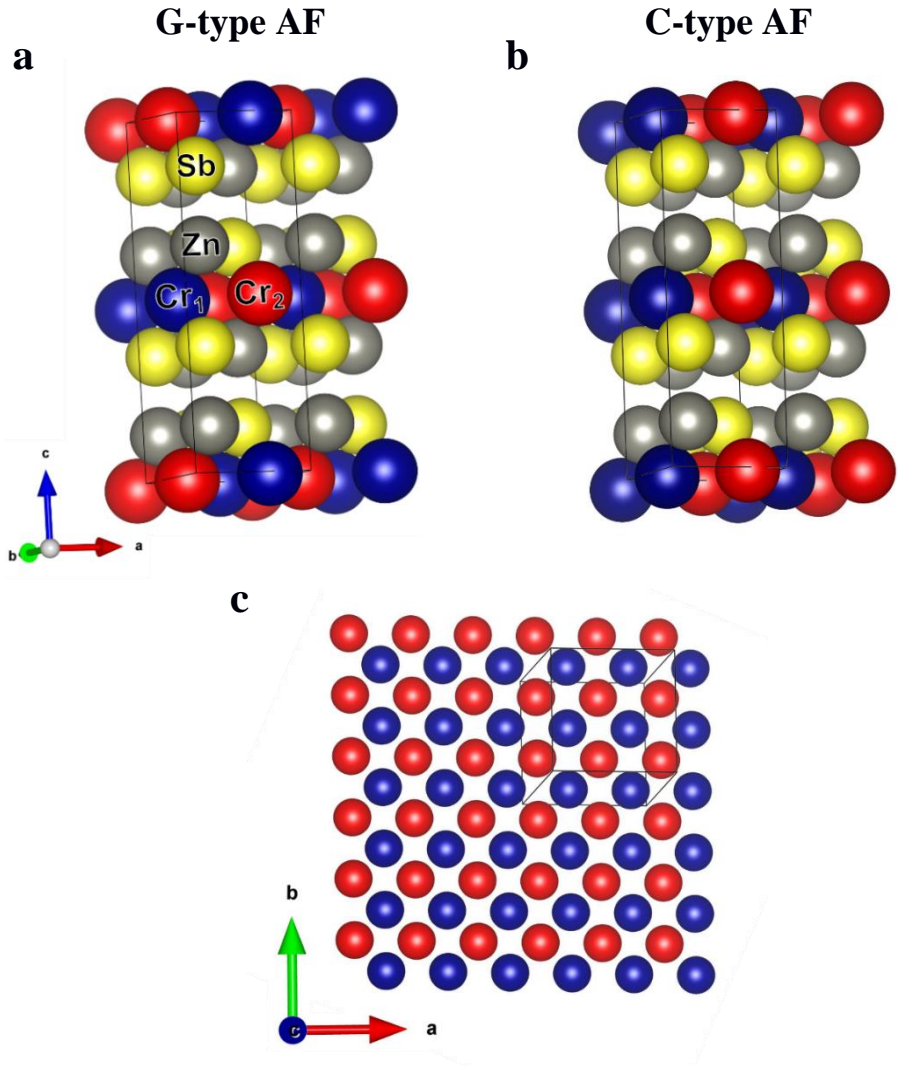

Figure. 2: Panel a, b): Sketch of the tetragonal MnAlGe *P4/nmm* structure of Cr$_2$Zn$_2$Sb$_2$ with c/a = 1.492 doubled in c-direction (plotted by VESTA [50]) in a) G-type AF and b) C-type AF magnetic ordering. Panel c): Cr-atoms in the ($\vec{a}, \vec{b}$) plane with checkerboard antiferromagnetic ordering [11]. Red and blue spheres refer to spin up/down Cr magnetic moments.

Table 2 lists the DFT derived atomic coordinates of the C-AF structure revealing the peculiar, layered properties of the *P4/nmm* structure: a) Cr atoms in ($\vec{a}, \vec{b}$) planes, which are sandwiched between mixed Zn-Sb planes, b) Cr atoms in chains in $\vec{c}$ direction alternating with Zn-Sb chains, and c) empty layers without Cr atoms. The same holds for the G-AF structure, for which there is now antiparallel magnetic moment ordering in the $\vec{c}$ direction.

For the G-AF spin configuration Figure 2a sketches the unit cell of $Cr_4Zn_4Sb_4$ (C-AF, doubled in c-direction). For G-AF in $\vec{c}$ direction the checkerboard layers are repeated with reversed magnetic moments. The energy difference between C-AF and G-AF is rather small, with G-AF being more stable by 0.59 kJ mol$^{-1}$($Cr_2Zn_2Sb_2$). Noticeable, for the G-AF structure in comparison to C-AF (see caption of Table 2) there is a contraction of the lattice parameter $c$ by 1.6 % and a stretching of $a$ by about 1.1%.

Table 2: DFT derived structural data for $Cr_2Zn_2Sb_2$ with *P4/nmm* structure and C-AF antiferromagnetic ordering. Lattice positions corresponding to the lattice parameters $a = 0.4206$ nm and $c = 0.6277$ nm. For $Cr_4Zn_4Sb_4$ with G-AF ordering the lattice positions are very similar but for comparison the z-coordinates have to be divided by 2 because of the doubling of the cell in c-direction. The second $Cr_2Zn_2Sb_2$ unit, which is ordered with antiparallel magnetic moments to the first unit, is constructed by adding (0,0,0.5) to the positions of the first unit. The lattice parameters for G-AF are $a = 0.4253$ nm and $c = 2 \times 0.6175$ nm.

| $Cr_2Zn_2Sb_2$ C-AF | | | |
|---|---|---|---|
| **Atom** | **x** | **y** | **z** |
| Cr1 | 0.75 | 0.25 | 0 |
| Cr2 | 0.25 | 0.75 | 0 |
| Zn1 | 0.25 | 0.25 | 0.273 |
| Zn2 | 0.75 | 0.75 | -0.273 |
| Sb1 | 0.25 | 0.25 | -0.287 |
| Sb2 | 0.75 | 0.25 | 0.287 |

To obtain a deeper understanding of the chemical bonding, also charge transfer was calculated for stoichiometric $Cr_2Zn_2Sb_2$ and is presented in the supplemental material [49].

### 3.4. DFT Vacancy structures of $Cr_{0.875}ZnSb$

As discussed in Chapter 3.1, the experimentally synthesized compound does not crystallize defect-free, rather the stable stoichiometry was found to be $Cr_{0.86}ZnSb$. Subsequently, structures with a stoichiometry of $Cr_{0.875}ZnSb$ were modelled, constructing supercells with composition $Cr_{14}Zn_{16}Sb_{16}$. The simplest ansatz for a Cr-vacancy structure with a composition close to experiment would be $Cr_7Zn_8Sb_8$, which, however, does not allow for perfect antiferromagnetic ordering, as the supercell contains an odd number of magnetic Cr atoms. To maintain a zero total magnetic moment for the modelled crystal, a 2×2×2 supercell containing 16 atoms per species was constructed, allowing to remove a suitable pair of antiparallel magnetic Cr atoms, resulting in $Cr_{14}Zn_{16}Sb_{16}$ with a zero total magnetic moment.

In general, the formation of Cr vacancies is supported by our DFT calculations, which result in an energy difference of 2.4 kJ/mol (CrZnSb), by which G-AF ordered $Cr_{14}Zn_{16}Sb_{16}$ + 2 Cr atoms arranged in a bcc structure is more favourable than vacancy-free $Cr_{16}Zn_{16}Sb_{16}$ with G-AF ordering. For the vacancy supercells, G-AF ordering was found to be more stable than the respective C-AF structures. The differences, however, are rather small, a fraction of 1 kJ mol$^{-1}$ ($Cr_{0.875}ZnSb$). For G-AF ordering, the two most stable atomic arrangements are depicted in Fig. 3. For *vac*1-2 G-AF in panel b) of Fig. 3 one observes rows with vacancies alternating with defect-free rows in the ($\vec{a}$, $\vec{b}$) plane. Along $\vec{c}$, this distribution is repeated but with up/down moments reversed. Panel a) of Fig. 3 illustrates the *vac*1-9 C-AF structure, with two neighbouring defect rows in the ($\vec{a}$, $\vec{b}$) plane. Along $\vec{c}$, an ($\vec{a}$, $\vec{b}$) plane with defect rows alternates with defect-free planes.

As the formation energies are rather similar, a DFT prediction of the most stable structure is not unambiguous. Interestingly, however, the two discussed structures exhibit distinctly different electronic DOS features close to Fermi energy, of which only *vac*1-2 G-AF satisfactorily reproduces the measured transport properties (see chapter 5).

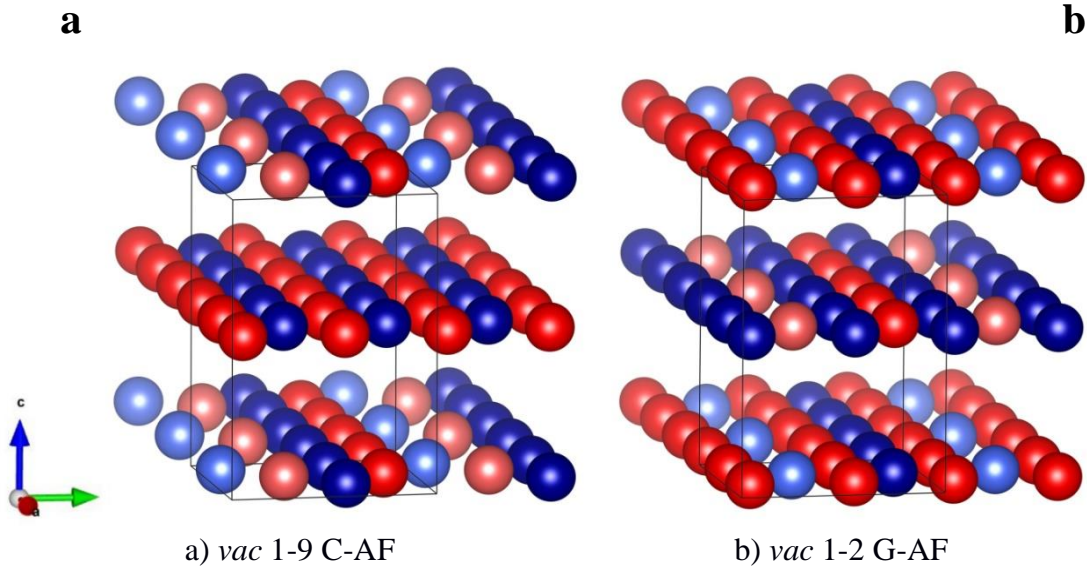

a) *vac* 1-9 C-AF          b) *vac* 1-2 G-AF

Figure 3: DFT structures with two Cr-vacancies per 46-atom supercell. Only spin up/spin down (red/blue) Cr atoms are shown. Panel a): *vac*1-9 structure with C-AF ordering. Panel b): *vac*1-2 structure with G-AF ordering. Atoms in vacancy rows appear in lighter colours.

For the *vac*1-2 G-AF structure with composition $Cr_{1.75}Zn_2Sb_2$ the DFT equilibrium volume is 0.1096 nm$^3$, which compares favourably with the slightly smaller experimental volume of 0.1082 nm$^3$ for $Cr_{1.72}Zn_2Sb_2$. The DFT lattice parameters of the supercell $a/2 = 0.4216$ nm, $c/2 = 0.6166$ nm are in good agreement with the experimental data of a = 0.418 nm, c = 0.619 nm. The DFT volume for $Cr_2Zn_2Sb_2$ without vacancies of 0.1115 nm$^3$, however, is larger than the vacancy value by about 1.7%. Switching on a DFT+U approach [35] for the Cr-d states increases the equilibrium volumes of the *vac*1-2 G-AF structure by 2.4% (U-J = 1 eV), 4.9% (U-J = 2 eV) and 7.4 % (U-J = 3 eV). Clearly, such an approach (even for a rather modest value of U-J = 1 eV) is not advisable for deriving a correct ground state volume.

Because the measured samples are synthesized at elevated temperatures, a suitable thermodynamic average of the relaxed supercell structures needs to be considered for a comparison of structural details. It can, however, be safely assumed that the small distortions of lattice coordinates upon vacancy formation - as they appear in the relaxed DFT calculations - average out. On the other hand, the average must ensure in some way the dominance of the pseudo-gap feature as seen in the DOS of *vac*1-2 G-AF, because of the agreement of measured and calculated electronic transport properties reported in section 5.

### 3.5. DFT Electronic structure of Cr$_{0.875}$ZnSb

All density of states (DOS) calculations were performed for antiferromagnetic ordering. All the total spin up and down DOS, summed over all atoms of the Cr-, Zn- and Sb sublattices, are the same for both directions. The local atomic DOS contributions are the sum over all the ℓ-projected DOS contributions within a sphere surrounding each species of atoms (see Section 2.2). The spin up DOS for the *vac*1-2 G-AF structure in Fig. 4a shows distinct features below -3 eV, where Zn s-like and Sb p-like states dominate. Then, hybridization with Cr states sets in and above -2 eV the DOS is dominated by Cr d-like states up to 2 eV.

A striking pseudo-gap-like minimum forms right above the Fermi level at $E_{gap}$ = 0.12 eV. A closer look (see Fig 4b) illustrates the rapid fall and rise of the DOS above and below $E_{gap}$. If $E_F$ is shifted up to $E_{gap}$ then two electrons (one spin up/down) need to be doped to the vacancy system. This gap-like feature occurs, even when the structure is not relaxed; therefore, it cannot be explained by structural relaxation as caused by the vacancies. It also exists, if perfect antiferromagnetic ordering is destroyed, as revealed by a DFT calculation for a Cr$_7$Zn$_8$Sb$_8$ vacancy supercell (Fig. S1, Suppl. Mater.). For that, the unit cell of perfect Cr$_2$Zn$_2$Sb$_2$ was doubled in the $(\vec{a},\vec{b})$ plane resulting in Cr$_8$Zn$_8$Sb$_8$, which by removing one spin-up Cr atom ends up in Cr$_7$Zn$_8$Sb$_8$. Because of the removed spin-up Cr atom, the total magnetic moment is now -3.47 $\mu_B$. This vacancy structure is energetically less stable by only 0.61 kJ mol$^{-1}$ (Cr$_{0.875}$ZnSb) than the perfect antiferromagnetic *vac*1-2 AFB structure. Inspecting the DOS, one detects again a similar deep gap-like feature at -62 meV now below $E_F$ (Fig. S1, Suppl. Mater.).

In Fig. 4b, by comparing the DOS of the vacancy free Cr$_{16}$Zn$_{16}$Sb$_{16}$ structure with the pseudo gap case, the effect of the vacancies is strikingly illustrated: at the energy of the pseudo gap, a peak of the DOS of the vacancy free case arises. Figure 4c depicts the band structure of *vac*1-2 G-AF. The red and green dashed lines mark the positions of the two doping energies as discussed in section 5. Noticeably, at $E_F$ up to 0.15 eV the number of bands along Γ-X-M-Γ is strikingly reduced compared to energies below.

Figure 4d depicts the densities of states of a different vacancy structure *vac*1-9 for G-AF and C-AF, respectively. For this structure, no pseudogap is discernible, as the calculations reveal a metallic density of states close to $E_F$. Notably, this disagrees with the experimental results displayed in the subsequent sections.

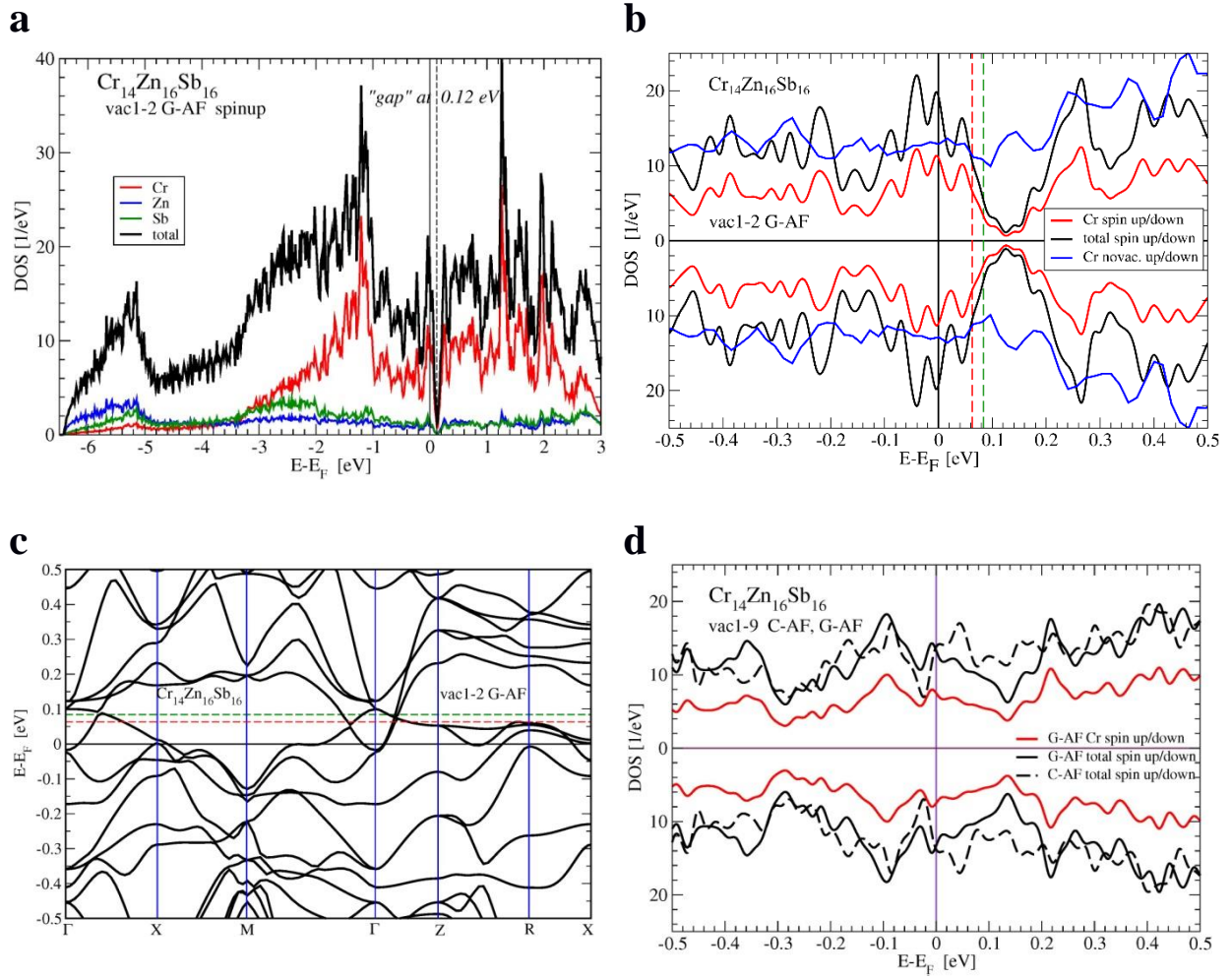

Figure 4: Panel a): Spin up DOS of *vac*1-2 G-AF. The DOS for only one spin direction is shown because of symmetry the DOS is the same for the second spin direction. Local densities of states summed over all ℓ for Cr (red), Zn (blue) and Sb (green). Panel b): DOS close to $E_F$ for G-AF ordered vac1-2 $Cr_{14}Zn_{16}Sb_{16}$ and stoichiometric CrZnSb. The black, red and blue lines refer to the total DOS, the local DOS for Cr, as well as to the total DOS for the vacancy-free G-AF structure, respectively. Panel c): Band structure of *vac*1-2 near the Fermi level. The dashed lines again refer to $E_F$ for the two doping levels. Panel d): DOS close to $E_F$ for *vac*1-9 for C-AF (dashed line) and G-AF (full lines) ordering.

Analysing the bonding of the states around $E_F$ for the perfect structure Cr$_{16}$Zn$_{16}$Sb$_{16}$, one finds Cr d-states with predominantly d$_{xz}$ and d$_{yz}$ character which interact with Sb p-states of p$_z$-character. The Cr-Zn and Cr-Sb bond lengths of 0.271 nm (Table 1b) are the shortest of the compound. Part of these p-d bonds are broken, when Cr-vacancies are introduced. This bond breaking also happens for all the Cr-vacancy structures, but an expressed gap-like feature in the DOS appears only for the *vac*1-2 G-AF structure. There, chains of Cr vacancies in $\vec{c}$-direction are formed. Further analysis shows also s-like and smaller p-like DOS features for Zn.

As elaborated in chapter 5, a distinct feature of the measurement data is the temperature-dependent Seebeck coefficient, with a maximum of 64 µV K$^{-1}$ at around 178 K. The electronic structure and DOS at $E_F$ cannot explain such a high Seebeck coefficient. Rather, the Fermi energy needs to be shifted closer to the gap-like feature by an electron doping of 1.46 $e^-$ for a composition of Cr$_{14}$Zn$_{16}$Sb$_{16}$, yielding reasonable agreement with our measurements (see section 5.2). Then, $E_F$ lies 62 meV above the undoped case.

### 3.6. Consideration of different defects

Inspecting the DOS of Figure 4b of the vac1-2 G-AF structure one finds the discussed minimum at exactly 2 electrons above the Fermi energy. An additional Zn atom with two valence electrons would place $E_F$ in the pseudogap, which is confirmed by the self-consistent calculation for one Zn atom placed at a Cr vacancy site, yielding a stoichiometry of Cr$_{14}$Zn$_{17}$Sb$_{16}$. Figure 5 shows the DOS of the fully relaxed structure. Evidently, $E_F$ is located right below the pseudogap regarding the spin up and down DOSs, which are different, because the corresponding symmetry, i.e. perfect antiferromagnetism, is destroyed. A small total magnetic moment of about 0.01 µ$_B$ results. Placing the Zn atom at the second Cr vacancy site instead, the spin dependent DOSs are reversed. Another effect of the shift of $E_F$ into the pseudogap is the stabilization of the vac 1-2 G-AF structure, which becomes more stable than the competing vac1-9 G-AF structure by 0.54 kJ.mol$^{-1}$ (Cr$_{0.875}$Zn$_{1.0625}$Sb). The magnitude and distribution of the site projected local moments of Cr are similar to the results for Cr$_{14}$Zn$_{16}$Sb$_{16}$.

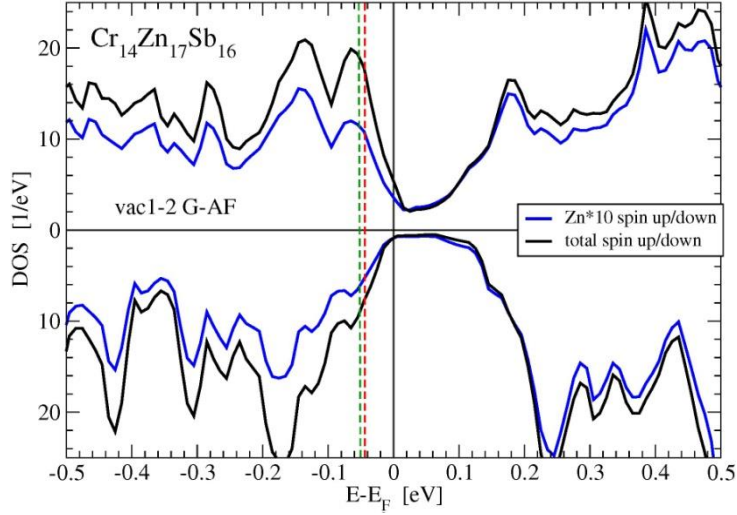

Figure 5: Electronic DOS close to $E_F$ of Cr$_{14}$Zn$_{17}$Sb$_{16}$. One Zn atom is placed at a spin up Cr vacancy site of the vac 1-2 G-AF structure. Red and green dashed vertical lines indicate the respective Fermi energies of positive doping by removing 0.6 and 0.76 electrons. The total spin polarized DOS (black line) and the corresponding local Zn DOS (blue line, summed over all Zn atoms), enlarged by a factor of 10, are shown.

Figure 5 also shows two more Fermi energies determined by positive doping (less electrons) of 0.6 and 0.76 electrons, cutting through the steep descending slope of the DOS next to the pseudogap. An adjustment of the Fermi energy is necessary to align the transport properties with experiments (see S5). The figure also highlights that the features of the local DOS of Zn near $E_F$ closely follow the total DOS, indicating hybridization of the states close to the pseudogap.

Another defect structure was studied for Cr$_{14}$Zn$_{15}$Sb$_{17}$ by substituting a Zn atom with Sb next to a Cr vacancy. The analysis of the DOS again displays the hallmark of the pseudogap which appears about 1 electron above $E_F$. For a doping of 0.5 electrons, the temperature-dependent Seebeck coefficient and resistivity were calculated and reveal similar behaviour to the experimental results. Because the DOS minimum is less steep in comparison with the previously discussed cases, the maximum of the positive Seebeck peak is significantly lower. Similar to Cr$_{14}$Zn$_{17}$Sb$_{16}$, the antiferromagnetism is not perfect anymore, because of the symmetry reduction and therefore a small total moment of -0.03 $\mu_B$ arises.

Based on the defect investigations, it is conceivable that the real system includes a mixture of defect structures. For example, a mixture of $(1-x)$(Cr$_{14}$Zn$_{15}$Sb$_{17}$) + $x$(Cr$_{14}$Zn$_{17}$Sb$_{16}$) would result in Cr$_{0.857}$ZnSb for $x=2/3$, which is very close to the experimental data.

## 4. Magnetic properties

Magnetic susceptibility and specific heat measurements of a nominally stoichiometric CrZnSb sample were reported by Ciesielski et al. [11]. Their magnetic susceptibility data revealed a distinct anomaly near 230 K, which however, was attributed to antiferromagnetic ordering of an unspecified impurity phase and the intrinsic magnetic behaviour of CrZnSb was concluded to be paramagnetic.

### 4.1. Magnetic susceptibility and magnetization studies

The temperature dependent dc magnetic susceptibility of Cr$_{0.86}$ZnSb, $\chi(T) \equiv M/H(T)$, measured at an applied field $\mu_0 H = 0.1$ T (Fig. 6), reveals an antiferromagnetic-like cusp at $T_N \sim 220$ K, which is superimposed to a rather large, weakly temperature dependent Pauli-like susceptibility, $\chi_0 \sim 10^{-3}$ emu/mol. Despite of slightly different nominal stoichiometries of the present Cr$_{0.86}$ZnSb sample as compared to earlier reported CrZnSb, we observe a very close match of the 0.1 T susceptibility data with those shown in Ref. [11] and we, thus, assign the revealed features of the magnetic susceptibility to an intrinsic behavior of the Cr$_{0.86}$ZnSb matrix phase. An itinerant, weak-antiferromagnetic nature of the magnetic state of Cr$_{0.86}$ZnSb is indicated by the large T-independent contribution $\chi_0$ and by a super-linear field dependence of the isothermal magnetization M(H), measured at low temperature ($T = 3.5$ K) and displayed as inset of Fig. 6. The latter is indicative of a broadened metamagnetic transition, occurring at fields up to 1 T. As revealed in Fig. 6 by a distinct difference between M(T)/H curves measured at 0.1 and 1 T, an initially non-linear field dependence of the magnetization persists for temperatures up to $T_N \sim 220$ K.

The temperature dependent heat capacity $C_p(T)$ of Cr$_{0.86}$ZnSb, as depicted in Figure S3 of the supplemental material [49], reveals a small anomaly around 220 K, which is in good agreement with an itinerant antiferromagnetic phase transition, aligning with the magnetic susceptibility data. From the low-temperature data of $C_p(T)$, we further determined a Sommerfeld coefficient of $\gamma = 10.7(5)$ mJ.mol$^{-1}$K$^{-2}$. DFT derived values of $\gamma$ are notably lower, indicating an effective mass enhancement due to spin fluctuations as detailed in the supplemental material [49].

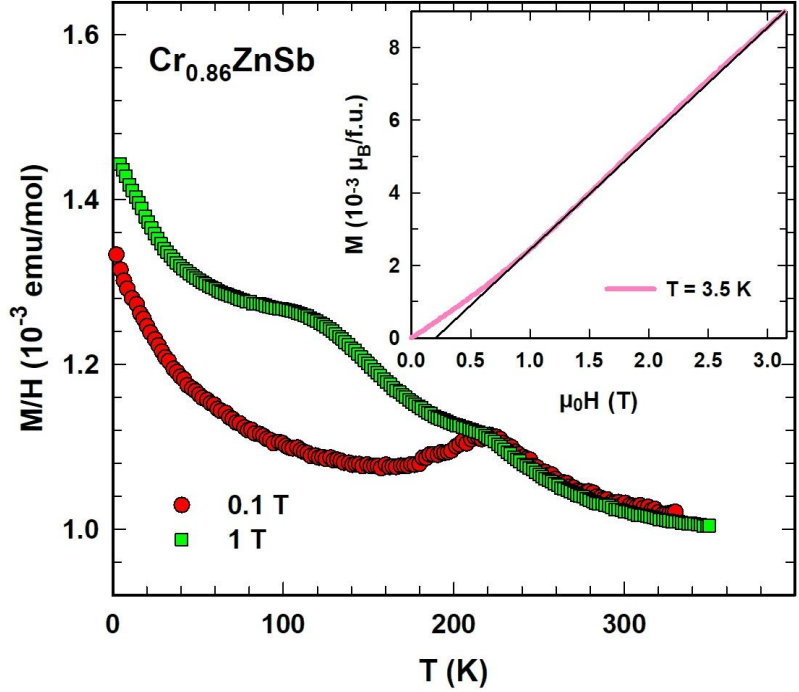

Figure 6: Temperature dependent magnetic susceptibility, *M/H*(T) of Cr$_{0.86}$ZnSb measured at applied fields as labeled. The inset shows isothermal magnetization data, M(H), measured at 3.5 K. The straight black solid line is a linear fit to *M* (H > 1 T) intended to uncover the non-linear field dependence of the isothermal magnetization at fields below 1 T.

### 4.3  Muon spin relaxation study of the magnetism in Cr$_{0.86}$ZnSb

The rather weak features of antiferromagnetic order revealed by the above magnetic susceptibility and heat capacity results may not be sufficiently conclusive with respect to an intrinsic bulk or extrinsic impurity origin. We, thus, complemented the macroscopic bulk measurements with microscopic local probe µSR studies. The latter were executed at 24 temperature steps in zero external magnetic field whereof seven selected relaxation data are displayed in Fig. 7. µSR data taken in the paramagnetic state of Cr$_{0.86}$ZnSb, i.e. at 300 K, refer to an initial $t = 0$ asymmetry (A$_{initial}$) near 27% which corresponds to the initially fully spin polarized state of the implanted muons. The dynamic relaxation due to paramagnetic fluctuations of the local magnetic fields at the muon stopping sites in Cr$_{0.86}$ZnSb are reasonably well accounted for by a simple Lorentzian-type exponential relaxation function. However, at temperatures below 240 K and in particular in the temperature interval 230 K to 210 K, dynamic relaxation transforms from a simple exponential to a stretched exponential behaviour, A(t) = A$_R$·exp(-λ$t^β$), on top of a non-relaxing component A$_{NR}$ ~ 12 %. The magnitude of the latter component is in realistic correspondence to the spot size of the muon beam and the density of the measured sample for being attributed to the fraction of muons stopping in the sample holder made from Ag metal which displays a negligible zero-field muon spin depolarization. In order to allow us to parameterise the dataset over the temperature range 20K to 300K, and to avoid correlations between parameters, the function $A(t) = A_R \cdot \exp(-\lambda t^\beta) + A_{NR}$ was fitted with $A_{NR}$ fixed at 12 % and with the stretching exponent fixed to a constant value $\beta = 0.5$. We note, that adjusting these parameters does not affect the essential conclusions discussed below. Obtained fits are shown as solid lines on top of the measured data in Fig. 7. The key feature revealed in Fig. 8 is a step-like temperature dependent evolution of the extrapolated initial asymmetry, $A_{initial}(T) = A_R(T) + A_{NR}$, at around $T_N \sim 220$ K with a truly significant reduction of the extrapolated initial asymmetry $A_{initial}$ from about 27% towards 16%. This corresponds to a loss of roughly two-third of the initial signal originating from the sample. Such step-like evolution of $A_0(T)$ is typical for a state of long-range magnetic order which causes large enough quasi-static magnetic fields at the muon stopping sites to depolarize muons within the initial detector dead-time of 0.15 µs following muon implantation at the EMU instrument [28]. The observed dynamic relaxation at $t > 0.15$ µs is attributed to those muon stopping sites where the incoming muon spins are oriented parallel to the local magnetic field and thus remain unaffected by the local magnetic field unless that local field displays transverse magnetic fluctuations. The latter is obviously strongest right below the Néel temperature and slows down at temperatures well below $T_N$ (compare the temperature dependent evolution of relaxation rate $\lambda(T)$ as depicted in Fig. 8). Accordingly, dynamic relaxation effects become almost invisible at 20 K, where $A(t) \sim 16\%$ appears nearly time independent. The strong short time depolarization effect, which becomes a truly substantial bulk effect at around the Néel temperature $T_N \sim 220$ K, provides clear evidence for the occurrence of intrinsic antiferromagnetic order in Cr$_{0.86}$ZnSb.

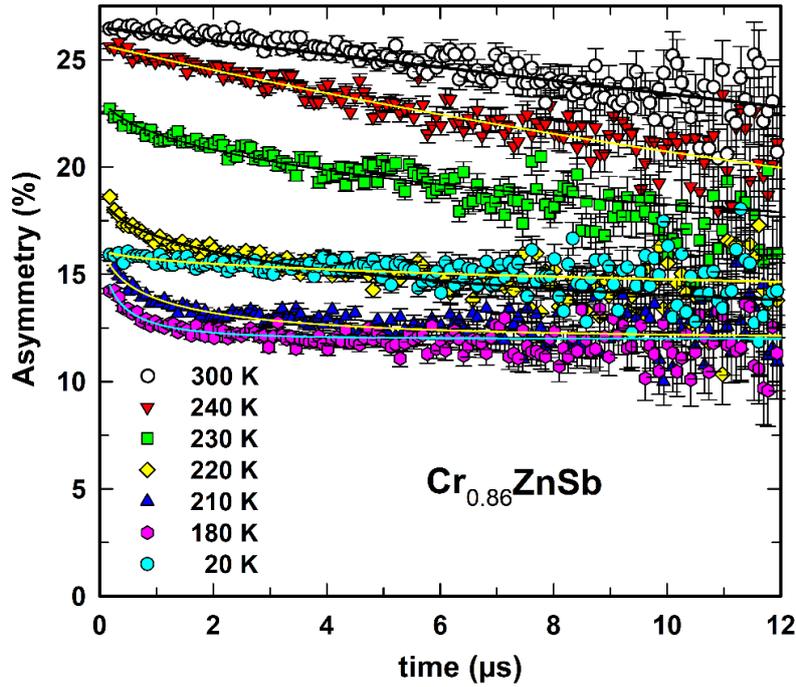

Figure 7: The time dependent zero-field µSR asymmetry data of $Cr_{0.86}ZnSb$ at selected temperatures as labeled, with fits shown as solid lines (see text).

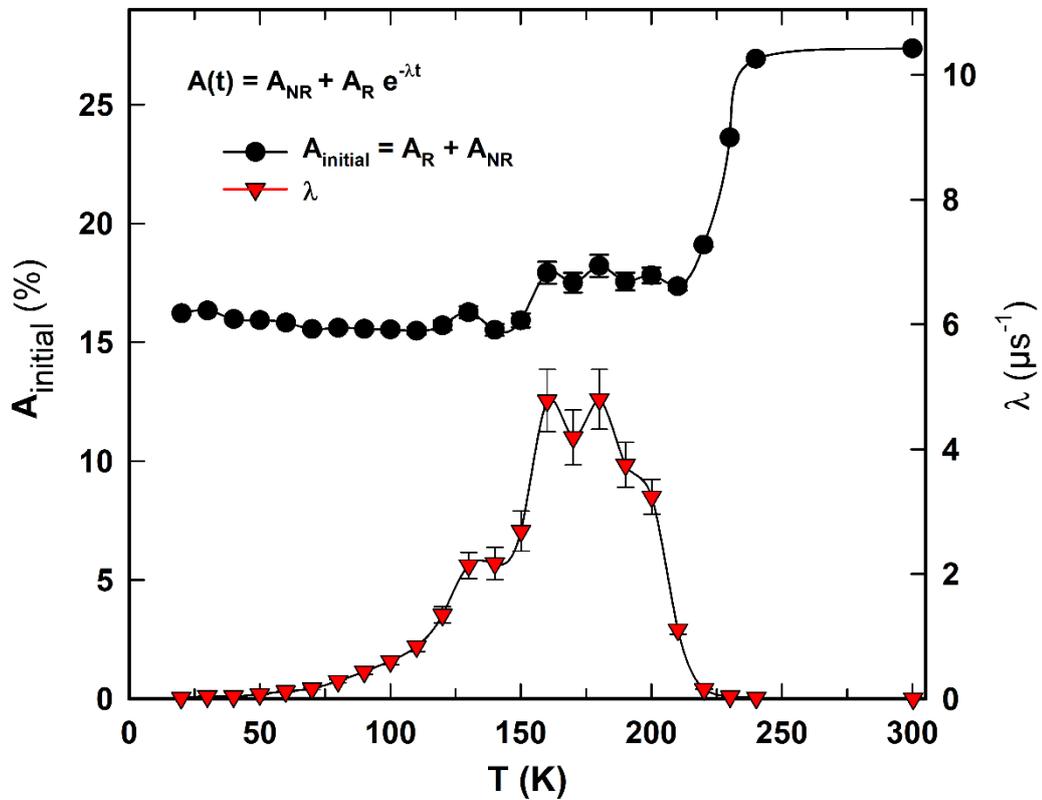

Figure 8: The temperature-dependent extrapolated initial asymmetry $A_{initial} = A_R + A_{NR}$ and muon spin depolarization rate $\lambda$ of $Cr_{0.86}ZnSb$ resulting from fits with a stretched exponential function with $A_{NR} = 12\%$ and stretching exponent $\beta = 0.5$ being fixed in a batch fit.

### 4.4 DFT-derived magnetic properties

The DFT calculation for the non-spin-polarised compound Cr$_2$Zn$_2$Sb$_2$ with relaxed *P*4/*nmm* structure results in a large value of the DOS at $E_F$ as also shown in Ref. [11]. For that, Cr-d-like states are mainly responsible as reflected by the local DOS of Cr at $E_F$, N(Cr) ~ 2 eV$^{-1}$. On the basis of Stoner's instability criterion for spontaneous ferromagnetic spin polarisation, $IN_{Cr}(E_F)$ > 1 with an assumed exchange energy for Cr of I ~ 0.8 eV it is obvious that the electronic structure of the non-spin polarised compound Cr$_2$Zn$_2$Sb$_2$ is unstable against ferromagnetic spin polarisation. The DFT ground state density is spin polarised and, based on the site projected spin densities, local magnetic moments are defined at the respective atomic sites. Therefore, comparison with experiment is only meaningful when results of spin polarised DFT calculations are considered.

Although a ferromagnetic (FM) ground state of Cr$_2$Zn$_2$Sb$_2$ is found, which is more stable by 33 kJmol$^{-1}$ (CrZnSb) than the non-polarised result, antiferromagnetic magnetic orderings (AF) need to be studied because of the presence of Cr. Investigating bcc Cr at the calculated equilibrium volume, small magnetic moments with antiferromagnetic ordering in a simple cubic lattice are established, in agreement with the results of [51]. There it is mentioned that the energetical proximity of FM and AF ordering may indicate the formation of a spin-density wave (SDW). According to the present DFT calculations, the energy difference between Cr$_2$Zn$_2$Sb$_2$ C-AF and the corresponding FM phase is approx. 4.8 kJ mol$^{-1}$(CrZnSb). Formation of SDW cannot be excluded a priori. It is well known that many Cr-alloys and compounds reveal SDWs [52]. Also, simple cubic Cr$_2$ is prone to formation of SDWs below 110 K, which is due to Fermi surface nesting features [51, 53]. Nevertheless, the magnetic experimental results of the present work do not indicate any formation of SDW.

The AF backbone of Cr$_x$ZnSb is the planar checkerboard arrangement as illustrated in Fig. 2b. For the DFT derived planar lattice parameter of a = 0.4206 nm for Cr$_2$Zn$_2$Sb$_2$ C-AF the 4 nearest neighbour distances are stretched by a factor 1.2 in comparison to the 8 nearest neighbour distances of the simple cubic AF ground state structure. Performing a DFT calculation for the stretched simple cubic structure results in large spin up/down local moments of ±3.1 µ$_B$ with a very small DOS at $E_F$. Doing the same for FM ordering with starting local moments of 2 µ$_B$ leads to zero magnetic moments after self-consistency has been achieved, the spin polarisation breaks down. It should be noted that the sizes and up/down orientations of the self-consistent moments are independent of the starting guess. Consequently, the total energy difference FM non-pol-AF is therefore very large, being greater than 90 kJ mol$^{-1}$ (Cr). Returning to the AF checkerboard structure, for the layer of pure Cr the DFT calculation predicts a semiconducting antiferromagnetic insulator with a gap of ~0.1 eV and large local spin-up/down moments of ~3.6 $\mu_B$ as well as total spin/up down moments of ±6 $\mu_B$, the total magnetic moment being zero (Suppl. Mat. Figs. S2a,b). A corresponding calculation for the FM case results in local magnetic moments of 3.8 $\mu_B$, very similar to the AF result, and a total magnetic moment of 5 $\mu_B$. The AF ordering is now more stable than FM by 32 kJmol$^{-1}$(Cr). A very similar energy difference is derived for Cr$_{14}$, with the *vac*1-2 G-AF structure, when all Zn and Sb atoms are removed. Now for the complete Cr$_{14}$Zn$_{16}$Sb$_{16}$ vacancy structure *vac*1-2 G-AF, the AF-FM total energy difference is strongly reduced to 8.3 kJmol$^{-1}$ (Cr$_{0.875}$ZnSb). For AF ordering, site projected local magnetic moments of Cr are in the range of 2.0 to 2.6 $\mu_B$, and for FM in the range of 2.0 to 2.2 $\mu_B$. Now, the interactions of Cr-3d states with the delocalized Sb-5p states have a significant impact. As listed in Table1 for the *P*4/*nmm* structure, Sb - as well as Zn - has 4 nearest-neighbour Cr atoms. Ideally, Sb is positioned above the center of a square formed by the 4 Cr atoms (if the Cr site is not vacant). They are ordered in two pairs of antiferromagnetically arranged Cr atoms. As a consequence, the induced site projected local magnetic moments in the Sb atomic sphere are zero. This is different for the FM case, for which Sb magnetic moments of 0.08 $\mu_B$ are formed, all of them antiparallel to the parallel moments of the ferromagnetic Cr moments. They are of purely p-orbital like character. For the delocalized Sb 5p-states, the field of Cr magnetic moments induces the formation of very weak magnetic moments of opposite orientation. Their delocalised character is also emphasised by the very small local DOS of the Sb 5-p states. Furthermore, according to the charge transfer analysis of section 3.3, Sb gains electronic charge, and consequently, in a simple model its atomic potentials gets less attractive and the atomic states are more delocalized.

For Cr$_{14}$Zn$_{16}$Sb$_{16}$ one might distinguish three different magnetic features. The delocalised Sb-5p states form some kind of polarisable electronic cloud between the Cr atoms screening the Cr moments. They seem to respond in a paramagnetic manner to a magnetic field as discussed. Furthermore, a weak antiferromagnetic coupling in $\vec{c}$ direction (G-AF ordering) of about 0.3 kJ mol$^{-1}$(CrZnSb) is predicted between the antiferromagnetic Cr layers. Finally, the site projected local magnetic moments of Cr atoms are large, in the range of 2 - 2.6 $\mu_B$ in the chosen atomic sphere of Cr, and they couple antiferromagnetically in the ($\vec{a},\vec{b}$) plane. These features seem to be reflected by susceptibility measurements (Fig. 6; see also Fig. S11a of Ciesielski et al. [22]). The peaks in the susceptibility for T > 200 K are attributed to magnetic order/disorder transition of the AF ordered screened magnetic moments of Cr in the ($\vec{a},\vec{b}$) plane. The predicted very weak interaction in $\vec{c}$ direction according to G-AF ordering is indicated by a feature in the susceptibility at T ~ 10 K (as a peak in [22] and as a flattening in Fig. 6, red curve).

## 5. Transport properties

### 5.1. Measurements and modelling

Transport properties are important to assess the potential of materials for thermoelectric applications and, furthermore, can give valuable insights into the electronic structure around the Fermi energy. To this aim, the temperature-dependent Seebeck coefficient and electrical resistivity of Cr$_{0.86}$ZnSb was studied in a broad temperature range (4 - 800 K), as shown in Fig. 9. The Seebeck coefficient, plotted as red triangles, exhibits a distinct maximum around $T = 180$ K, with values up to $S = 65$ µV/K. The positive sign of $S(T)$ indicates holes as dominant charge carriers up to high temperatures.

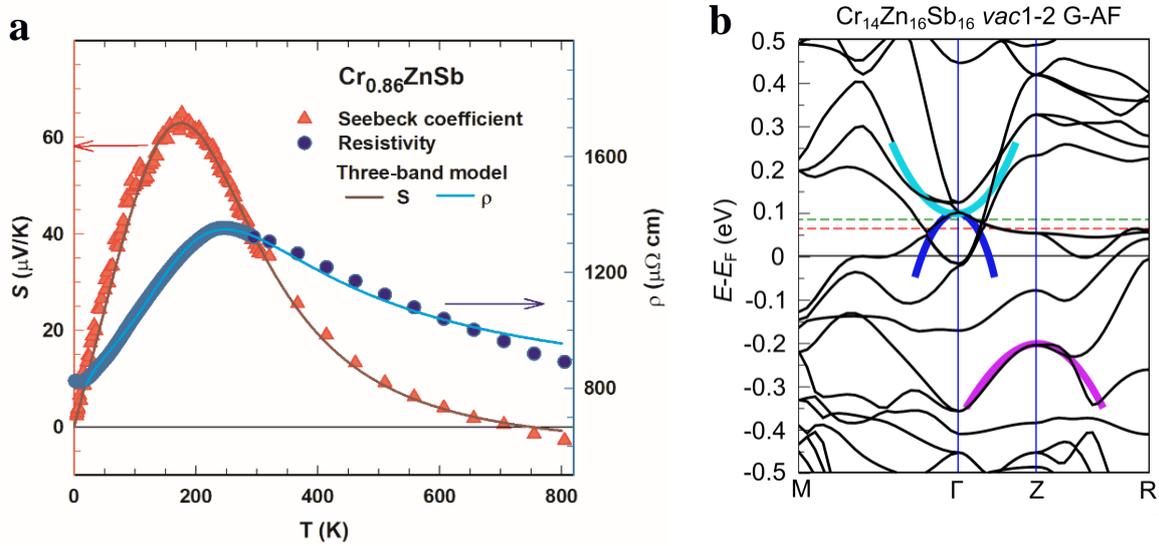

Figure 9: Panel a): Temperature-dependent Seebeck coefficient and electrical resistivity of Cr$_{0.86}$ZnSb. Solid lines are least squares fits employing a parabolic three-band model. Panel b): DFT derived band structure of Cr$_{14}$Zn$_{16}$Sb$_{16}$, together with a sketch of the band structure (solid, colored lines), obtained from modeling the temperature-dependent Seebeck coefficient of Cr$_{0.86}$ZnSb, employing the three parabolic band model. The red and green dashed lines mark $E_F$ for the two doping levels of 1.46 e$^-$ and 1.75 e$^-$.

The existence of a distinct maximum, moreover, points towards a second band contributing to transport above this temperature. The sign change $S(T)$ at around 700 K marks the temperature, where this conduction band outweighs the valence band contribution to the Seebeck coefficient. The resistivity, plotted as blue circles, shows typical features of a heavily doped semiconductor or semimetal. At very low temperatures $\rho(T)$ saturates to a constant value and rises with rising temperature. At around 250 K a maximum evolves, with negative $d\rho/dT$ for elevated temperatures above, presumably due to an increase of the charge carrier concentration $n$ with temperature. To deepen the understanding of the transport behavior of Cr$_{0.86}$ZnSb, the temperature-dependent Seebeck coefficient and electrical resistivity was calculated within the framework of Boltzmann's transport theory, using a three-parabolic-band model. The model was fitted to the experimental data using a least-squares fit method in a two-step procedure [54]. The fitting results are depicted as solid lines in Fig. 9 (left panel). The measurement of the electronic structure around the Fermi energy (energy-independent terms cancel in the denominator and numerator of the Boltzmann expression). Assuming acoustic-phonon scattering as the dominant scattering mechanism, the parabolic band model yields for the Seebeck coefficient of a single band [55] (equ. 2):

$$S(T) = \frac{k_B}{e}\left[\frac{2F_1(\eta,T)}{F_0(\eta,T)} - \eta\right], \qquad (2)$$

where $F_i$ are the Fermi integrals of i$^{th}$ order and $\eta = \mu/k_B T$ is the reduced chemical potential. The electron relaxation time and its energy dependence can be approximated by a power-law relation $\tau \propto N^{-1}(E) = \tau_0 E^{-1/2}$ for acoustic-phonon scattering; here, $N(E)$ is the DOS, which follows $\sim E^{1/2}$ for parabolic bands and $\tau_0$ is an energy-independent prefactor. To obtain the total Seebeck coefficient for multiple bands (equ. 3), the single contributions must be weighted with the electrical conductivities of the respective bands, neglecting interband scattering:

$$S_{tot} = \frac{\sum_{i=1}^{N} S_i \sigma_i}{\sum_{i=1}^{N} \sigma_i}. \qquad (3)$$

As demonstrated successfully in studies on different materials, such as Fe$_2$VAl-based Heusler compounds [56-58] and skutterudites [59], band-specific parameters, such as the size of the band gap/overlap, relative band masses or the position of the Fermi energy can be extracted by modelling $S(T)$ (see Fig. 9). To validate our model results, we additionally fitted the experimental resistivity data in a broad temperature range, using the model parameters obtained from the Seebeck coefficient fit. In this step, only one single free parameter (constant prefactor) must be included, if the dominant scattering process is set *a priori*. At low temperatures, scattering from (ionized) impurities often dominates $\rho(T)$, while phonon scattering becomes more important at higher temperatures due to the strong temperature dependence of the scattering rate $\sim T^{3/2}$. Therefore, we accounted for two different scattering processes via Mathiessen's rule (ionized impurities + phonons), which yields remarkable agreement with experimental resistivity data, when the same band parameters from modelling the Seebeck coefficient are used. The fit results yield a semi-metallic band configuration, which best describes the measured transport properties. Here the Fermi level is doped significantly into the first valence band by about 48 meV, while the conduction band overlaps with that band, yielding a tiny pseudo-gap of around -2 meV. The observable negative curvature of the measurement data at elevated temperatures ($T > 300$ K) is modelled by a second valence band, situated further from the Fermi level at -326 meV. For comparison, the effective electronic structure obtained from our fit is illustrated together with the electronic band structure from DFT in Fig. 9 (right panel). It is important to note that only relative band masses are obtained from the fit, therefore only the relative curvature of the fitted bands is significant. Good agreement with DFT results for Cr$_{0.875}$ZnSb is found, when directly comparing the band structure with the assessed bands from the fit. The position of the Fermi level agrees very well with the shifted Fermi level obtained from the DFT calculations. Moreover, the position of the two valence bands aligns very well with theoretical bands. The fitted band overlap at $E_F$ is somewhat smaller than the overlap indicated by DFT, which is understandable considering the multiple conduction bands near the Fermi level. In our model, which includes only three bands, the two conduction bands at the Γ point are represented by a single band located at their combined center of mass. Naturally, the DFT result comprises a large number of bands, not all of which can be considered, when fitting the effective band structure from transport properties. In summary, our analysis of transport properties strongly suggests the presence of a pseudo-gap in the electronic structure, corroborating the findings from DFT calculations (see Figs. 4c,d). Good agreement is found, when directly comparing the fitted bands with the DFT calculation close to the Fermi level, corroborating the emergence of a pseudo-gap in off-stoichiometric Cr$_{0.86}$ZnSb. It is remarkable that the Cr vacancies lead to such a drastic change in the electronic density of states.

To further validate the findings, resistivity and Hall effect of the annealed sample were measured as a function of the magnetic field from 0-9 T in the temperature-range from 4 - 300 K. As observable in Fig. 10(a), the Hall resistivity $\rho_{yx}$ is linearly proportional to the applied field and decreases with increasing temperature. In the single-band model, the carrier concentration $n$ is proportional to $1/eR_H$, where $R_H$ is the Hall coefficient derived from the slope of the Hall resistivity. In Fig. 10(b), this value is plotted as a function of temperature. A significant rise of the carrier concentration occurs above 150 K, indicating the contribution of a second band around this temperature. This coincides with the maximum of the Seebeck coefficient $S_{max}$ shown in Fig.9(a) and marked here with a dashed green line at around $T = 180$ K. Clearly, this strong increase of the carrier concentration aligning with $S_{max}$ further confirms of the existence of a pseudo-gap in this system. The Hall mobility, $\mu_H$, (Fig. 10(b), left axis) decreases in an almost linear manner from 4 to 300 K. Notably, the sample exhibits negligible magnetoresistance, MR < 1 %, over the whole temperature range, which is plotted in Fig. S4 of the supplemental material [49].

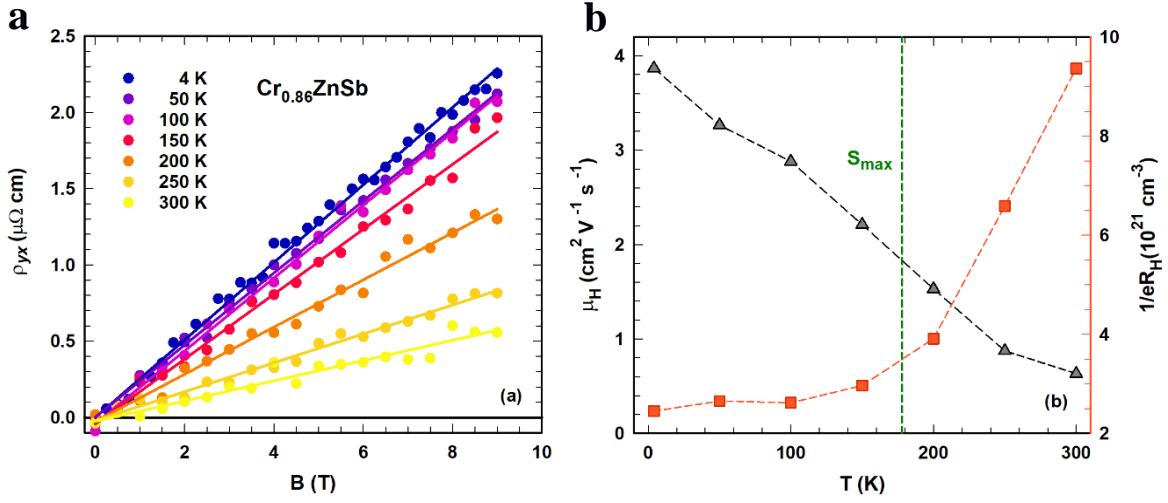

Figure 10: Panel a): Field-dependent Hall resistivity of Cr$_{0.86}$ZnSb for various temperatures. Panel b): Hall mobility µ$_H$ and $1/eR_H \propto n$ of Cr$_{0.86}$ZnSb. A clear rise of the inverse Hall resistance is found around the Seebeck maximum temperature $S_{max}$ indicating the contribution of a second band to electronic transport around this temperature, consistent with measurements of $S(T)$.

### 5.2. DFT Electronic Transport

Electronic transport calculations from DFT require the Kohn-Sham eigenvalues $E(\vec{k})$ on a sufficiently dense grid of $\vec{k}$ points in the Brillouin zone. Band velocities $\vec{v}(\vec{k}) = \frac{\partial E(k)}{\partial \vec{k}}$, are derived, which appear in integrals over the Brillouin zone including the energy derivative of the Fermi-Dirac distribution (equ. 4)

$$f^{FD}(E(\vec{k}),T,\mu) = \left((exp)\frac{(E(\vec{k})-\mu)}{k_B T}+1\right)^{-1}. \qquad (4)$$

The Fermi energy $E_F$ at elevated temperatures is the chemical potential $\mu(T)$ for the given number of valence electrons. Boltzmann's transport theory requires the integrals $K_{n,ij}$, i, j = 1, 2, 3 over the Brillouin zone of the tensor product of band velocities (equ. 5)

$$K_{n,ij} = \frac{1}{4\pi^3 \hbar} \int \tau\left(E(\vec{k})\right) v_i(\vec{k}) * v_j(\vec{k}) \left(E(\vec{k})-\mu\right)^n \left(-\frac{\partial f^{FD}}{\partial E(\vec{k})}\right) d\vec{k} \qquad (5)$$

involving the relaxation time $\tau(E(\vec{k}))$.

Then the components of the electronic conductivity tensor are (equ. 6)

$$\sigma_{ij} = e^2 K_{0,ij}. \qquad (6)$$

The resistivity tensor $\rho_{ij} = \sigma_{ij}^{-1}$ is the inverse of the conductivity tensor. The relaxation time $\tau$ [31] is determined by adjusting the isotropic DFT value to the measured isotropic resistivity at a chosen temperature.

Within the Boltzmann approach, the Seebeck tensor is defined by equ. 7

$$S_{ij} = \frac{K_{1,ij}}{eTK_{0,ij}} \qquad (7)$$

$e$ is the charge of an electron. In this expression, the constant relaxation time $\tau$ cancels out. Considering a tetragonal crystal symmetry - as is the case in the present study - the resistivity tensor and Seebeck tensors are diagonal and have three nonzero diagonal components $\rho_{ij} = \rho_{22}, \rho_{33}$ and $S_{11} = S_{22}, S_{33}$. The isotropic quantities $\rho$, $S$ are the averages of the three diagonal elements.

As shown in Fig. 11, panel a), the measured Seebeck coefficient $S_{exp}$ is positive up to temperatures of 700 K with a pronounced maximum of $S_{exp}$ (178 K) = 64 $\mu$V/K. In comparison the DFT result $S_{DFT}$ for the undoped case shows a broad positive peak with a maximum of ~ 40 $\mu$V/K at ~ 400 K. It always remains positive even at high temperatures. Clearly, the agreement with $S_{exp}$ is very poor, only the positive signs agree. According to Eqs. 2 and 3, $S$ is positive when states with energies $E(\vec{k})$ lower than $\mu$ dominate. Panel b) of Fig. 11 shows that for the undoped case $E_F$ lies in an energy region where the DOS $N(E_F)$ has a weak negative slope $N'(E_F)$. According to the low temperature approximation $S \propto -\frac{N'(E_F)}{N(E_F)}$, the sign of $S$ is then positive, which agrees with measurement, however, $S$ is too small, because the ratio is too small. The situation changes when 1.46 electrons are doped to the system. This value was chosen for achieving good agreement with experiment. Then, at the corresponding Fermi energy (red dashed line, Fig. 4b), the slope gets more negative, whereas the DOS decreases. Accordingly, $S$ should be larger than for the undoped case. The result of the full DFT calculation (red line panel a, Fig. 11) corroborates this argument: $S_{DFT}$ shows now a large peak of $S$ (234 K) = 78 $\mu$V/K. Overall, the DFT result for this doping agrees reasonably well with $S_{exp}$, even more so when the zero-line crossings at the same temperature T = 704 K are taken into account. The tetragonal splitting of the Seebeck tensor components can be quite substantial. For example, for the maximum at T = 234 K it is $S_{11} = S_{22} = 95$ $\mu$V/K and $S_{33} = 42$ $\mu$V/K.

Increasing the electron doping to 1.75 electrons per supercell the Fermi energy is further increased (green dashed line, Fig. 4b cutting the steep DOS at a reduced value $N(E_F)$ and $-\frac{N'(E_F)}{N(E_F)}$ is expected to be larger than for the doping of 1.46 electrons per supercell. However, this is not the case for the corresponding $S_{DFT}$ (blue curve panel a, Fig. 11), which is slightly reduced in height and also narrowed: at elevated temperatures, electronic states close or above the pseudogap gain more influence.

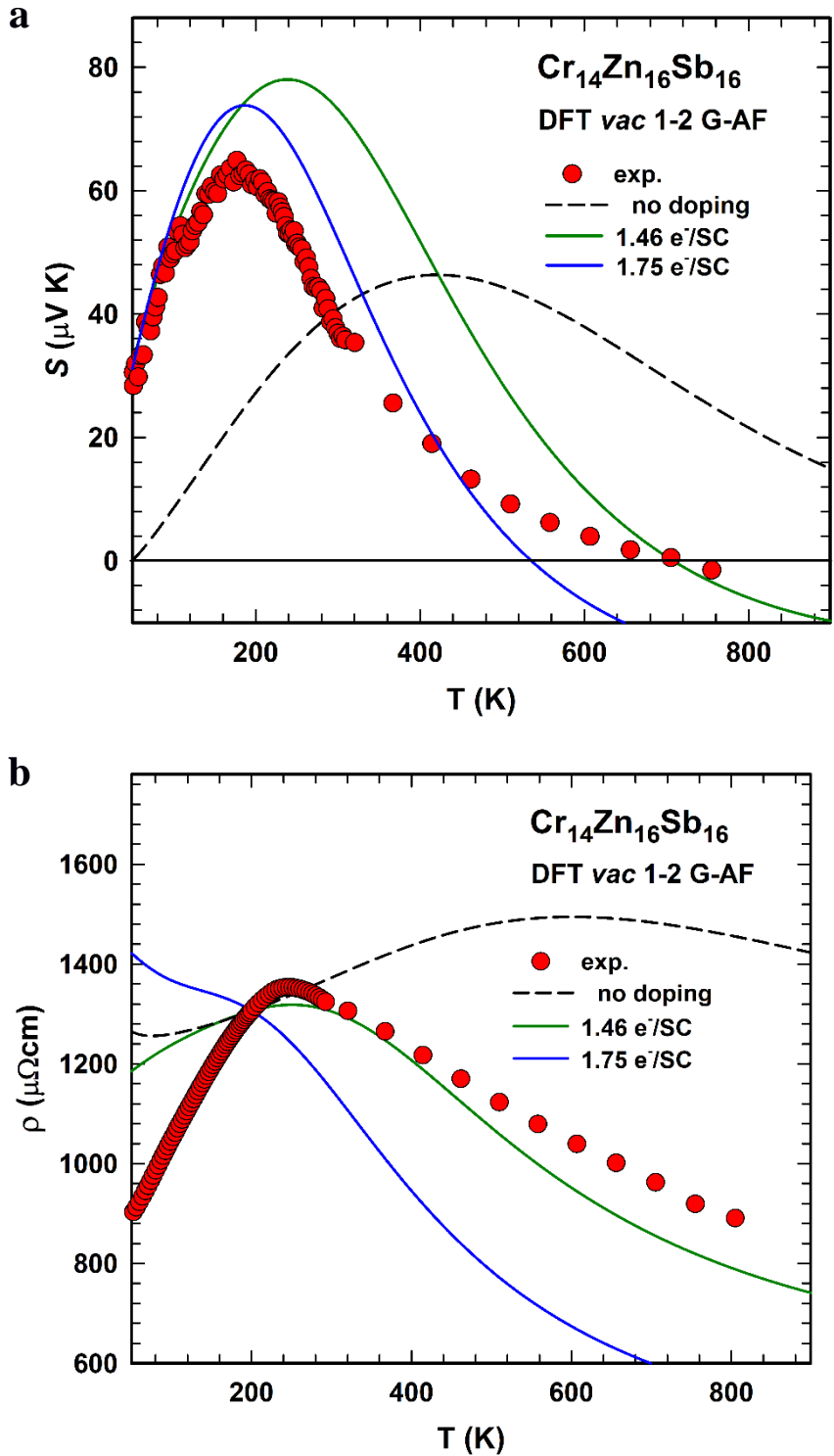

Figure 11: Measured and DFT calculated Seebeck coefficient and resistivity as a function of temperature for the *vac*1-2 G-AF structure of Cr$_{14}$Zn$_{16}$Sb$_{16}$. Measurements done for the sample Cr$_{0.86}$ZnSb. Panel a): Isotropic Seebeck coefficient $S$. DFT data for undoped case and two charge doping levels of -1.46 e$^-$/supercell and -1.75 e$^-$/supercell. Panel b): Isotropic resistivity $\rho$. DFT data for undoped case and two charge doping levels of 1.46 e$^-$/supercell. and 1.75 e$^-$/supercell. The relaxation time $\tau$ was determined according to $\rho_{DFT}$ (200 K) = $\rho_{exp}$ (200 K).

The crossing with the zero line is now at 536 K. Comparing just the two doped DFT Seebeck coefficients with measurement does not clearly indicate which one is better. However, the situation changes when resistivity results are considered.

Fig. 11b clearly shows a peaked behaviour of the measured resistivity $\rho_{exp}$. A maximum of 1354 µΩcm is reached at 240 K; subsequently $\rho_{exp}(T)$ decreases with rising temperatures down to 890 µΩcm at 806 K. As mentioned above, because of the unknown constant relaxation τ the DFT data for $\rho$ are fitted to the measurement at 200 K. Therefore, all curves cross at this temperature. The DFT resistivity for the undoped case is very different to $\rho_{exp}$, because it increases above 200 K up to a shallow maximum at 600 K. For the two doping levels, however, the DFT results change strikingly, showing also a decreasing $\rho$ with $T$ larger than 220 K. The decrease is most dramatic for the doping of 1.75 electrons, making the agreement with experiment rather poor.

However, for the doping of 1.46 electrons, the DFT $\rho$ above 220 K is just slightly smaller than measurement. A perfect agreement cannot be expected, because the $\rho_{DFT}$ does not include any effects of electron-phonon scattering, it is purely based on the band velocities of the electronic states. It is expected that at elevated temperatures electron-phonon scattering broadens the purely electronic $\rho_{DFT}(T)$ to a certain degree. Electron-phonon scattering is also neglected for $S_{DFT}(T)$. Presumably, the effects will cancel out to some extent, because of the fraction in equ.3. The mentioned doping levels are chosen with respect to the large supercell of composition Cr$_{14}$Zn$_{16}$Sb$_{16}$. By doping deviations of the stoichiometry of the measured sample from the ideal composition of the supercell are modelled assuming that - apart from shifting $E_F$ - the DFT electronic structure remains unchanged otherwise. With respect to the composition Cr$_{0.86}$ZnSb, the chosen doping amounts have to be divided by 16 assuming Zn or Sb are in excess, resulting in a stoichiometry of Zn$_{1.1}$ or Sb$_{1.1}$, its deviation from 1 being hardly measurable.

Discussing DFT $\rho$ below T = 200 K one observes three significantly different behaviours for the three different doping levels, with $\rho$(dop=-1.46 e$^-$/S.C.) as the most reasonable one, however, one should be careful about the interpretation of DFT $\rho$ and $S$ at very low temperatures. The reason is that the derivative of the Fermi-Dirac function $\frac{\partial f^{FD}}{\partial E(\vec{k})}$ is getting very narrow at very low temperatures. Therefore, for sampling a sufficiently large number of electronic states, the number of $\vec{k}$-points in the Brillouin zone must be very large. This can be done but it needs an extensive and time-consuming study which is beyond the scope of our paper.

## 6. Conclusion

Our comprehensive study of the $Cr_{0.86}ZnSb$ system through both experimental analysis and theoretical simulations has elucidated the unique properties of this compound arising from Cr vacancies. A detailed investigation of the $Cr_{1-x}ZnSb$ phase-space yielded only $Cr_{0.86}ZnSb$ as a stable, phase-pure compound, which crystallizes in the MnAlGe-type structure with Cr-vacancies. The experimental results obtained from magnetization and specific heat measurements reveal antiferromagnetic behavior with a Néel temperature of 220 K. In order to complement macroscopic bulk measurements with a local probe technique suited to reveal the approximate magnetically ordered volume fraction, muon spin relaxation measurements were conducted and the obtained temperature dependent zero-field relaxation data unambiguously confirm an intrinsic magnetic phase transition of $Cr_{0.86}ZnSb$ at around 220 K. Regarding the transport properties, measurements over a wide temperature range from 4-800 K were performed. Both, resistivity and Seebeck coefficients were found to exhibit distinct positive maxima in their temperature-dependent behavior. At about the same temperature, the inverse Hall coefficient exhibits a distinct upturn.

From analyzing the experimental data, the electronic structure was determined as a narrow-gap semimetal with a small pseudo-gap in the vicinity of the Fermi energy. Subsequently, DFT calculations for virtual $Cr_2Zn_2Sb_2$ as well as for vacancy structures were conducted using a supercell of composition $Cr_{14}Zn_{16}Sb_{16}$ to ensure perfect antiferromagnetic ordering. DFT confirms that the Cr vacancy structure is energetically more stable than the stoichiometric compound. Moreover, the analysis of formation energies regarding different spin configurations reveals $Cr_{0.875}ZnSb$ to be antiferromagnetic in line with experimental results. The calculations reveal large site projected local magnetic moments of Cr of 2-2.6 $\mu_B$ screened by delocalized p-states of Sb. The Cr vacancy defect structure with vacancies along the c-axis reveals a unique pseudo-gap in the DOS which drops down steeply at 2 electrons above Fermi energy. Such a feature is of importance for calculating suitable transport properties. By application of the BoltzTrap code [31], Seebeck coefficients and electronic resistivities are calculated directly from the DFT electronic structure, and excellent agreement with the measured data is achieved, when suitable doping is applied. The electronic structure of additional defects, such as placing Zn on a Cr vacancy site and substituting one Zn atom by Sb also shows the hallmark of the pseudo-gap.

Our research highlights the significance of structural defects in Cr$_{0.86}$ZnSb and formal consideration of vacancies and their configuration when analyzing such compounds. By combining theoretical methods with careful analysis of experimental data, we comprehensively characterized the magnetic and electronic structure, revealing the vacancy-driven formation of a pseudo-gap close to the Fermi energy in the antiferromagnetic Cr$_{0.86}$ZnSb.

**Acknowledgement**

This research was supported by MEYS, the Czech Republic via the bilateral mobility project 8J23AT014 co-sponsored by the Austrian WTZ-CZ08/2023 and by JST, project MIRAI. A part of this research was carried out at the EMU μSR beam line of the ISIS facility at the Rutherford Appleton Laboratory, UK. Beamtime was allocated via the Xpress proposal XB2291031.

# Supplementary information for

# Vacancy-induced pseudo-gap formation in antiferromagnetic Cr$_{0.86}$ZnSb

Michael Parzer[1*], Fabian Garmroudi[1], Herwig Michor[1], Xinlin Yan[1], Ernst Bauer[1], Gerda Rogl[2], Jiri Bursik[3], Stephen Cottrell[4], Raimund Podloucky[2], Peter Rogl[2]

[1] *Institute of Solid State Physics, TU Wien, Wiedner Hauptstr. 8-10, A-1040 Wien, Austria*
[2] *Institute of Materials Chemistry, Universität Wien, Währingerstr. 42, A-1090 Wien, Austria*
[3] *Institute of Physics of Materials, Czech Academy of Sciences, Žižkova 22, 61662 Brno, Czech Republic*
[4] *ISIS Facility, Rutherford Appleton Laboratory, Chilton, Didcot Oxon OX11 0QX, United Kingdom*

**Charge transfer**

To obtain a deeper understanding of the chemical bonding, charge transfer was calculated for stoichiometric Cr$_2$Zn$_2$Sb$_2$. Atomic Bader charges $q_B^{scf}$ were calculated for the self-consistently derived charge density, as well as the Bader charges $q_B$ for the superposed atomic charge densities. The difference $\Delta q_B = q_B^{scf} - q_B$ is defined as charge transfer for each atom and listed in Table 3. The charge transfer $\Delta q_B$ reflects the change of charge density after chemical bonding has been established. A similar definition was used for the volume change $\Delta V_B = V_B^{scf} - V_B$.

Table S1: Bader charge analysis of Cr$_2$Zn$_2$Sb$_2$ for the DFT-PBE calculation with the C-AF antiferromagnetic ordering. Charge transfers $\Delta q_B$ (in units of the proton charge) and volume changes $\Delta V_B$ (in 10³nm³, see text). $R_B^{scf}$ in nm of a sphere with volume $V_B^{scf}$.

|              | Cr    | Zn    | Sb    |
|--------------|-------|-------|-------|
| $\Delta q_B$ | 0.14  | 0.22  | -0.36 |
| $\Delta V_B$ | -0.46 | -0.82 | 1.31  |
| $R_B^{scf}$  | 0.148 | 0.158 | 0.182 |

According to Table S1 the charge transfers are rather small, being largest for Sb, which gains 0.36 electronic charge at the cost of Cr and Zn. Therefore, any theoretical model which is based on the assumption of larger ionicities is not adequate. As expected, the Bader volumes for Cr and Zn shrink, because both transfer electronic charge to Sb, for which the Bader volume is accordingly enlarged.

**Results from magnetic relaxation with different spin configurations of the Cr atoms (DFT)**

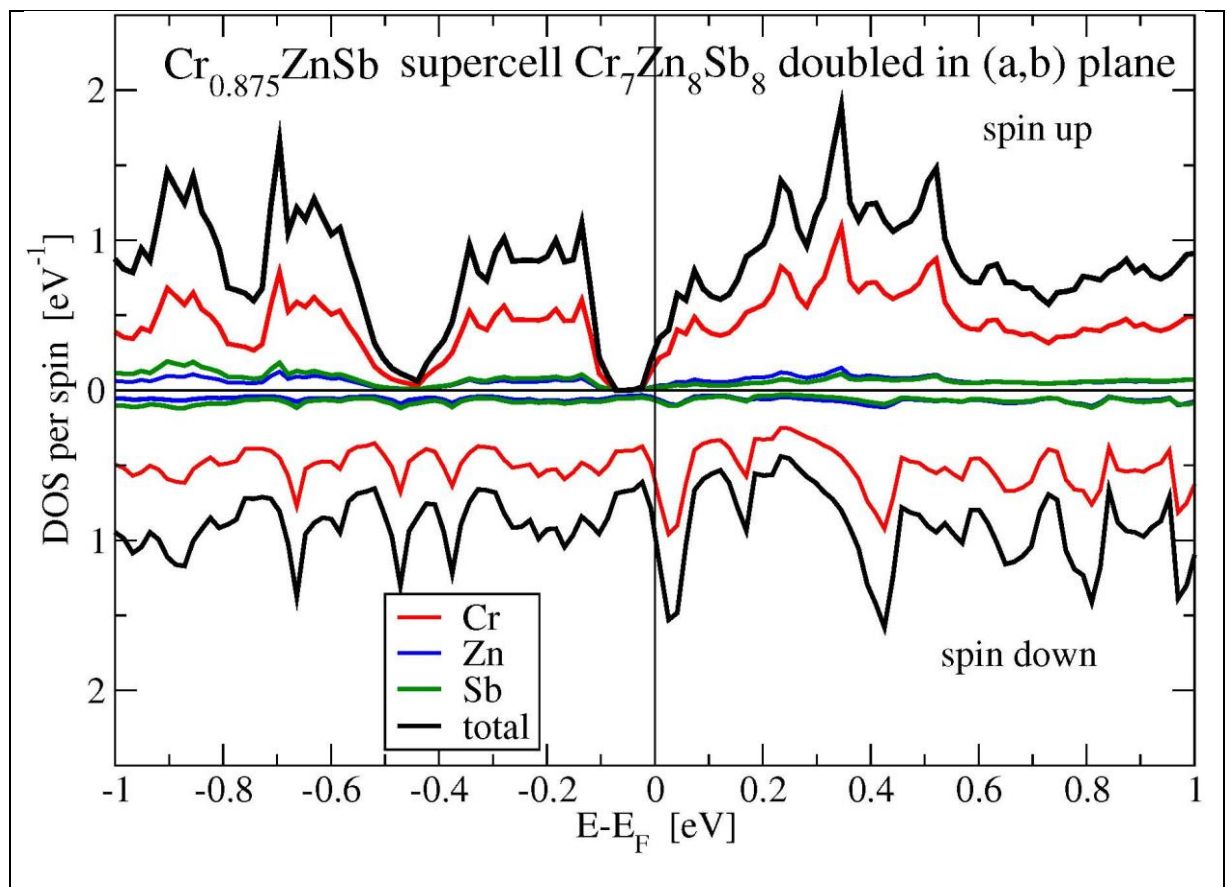

Figure S1: Total and local spin split DOS for $Cr_7Zn_8Sb_8$ as constructed by doubling the unit cell of *P*4*/nmm* $Cr_2Zn_2Sb_2$ in the (a,b) plane. Because only one Cr with a spin-up local magnetic moment is removed, the perfect antiferromagnetism of $Cr_2Zn_2Sb_2$ is destroyed. Hence, the total magnetic moment amounts to -4 $\mu_B$. There is a gap in the spin-up DOS at -0.05 eV as well as a deep minimum of the spin-down DOS.

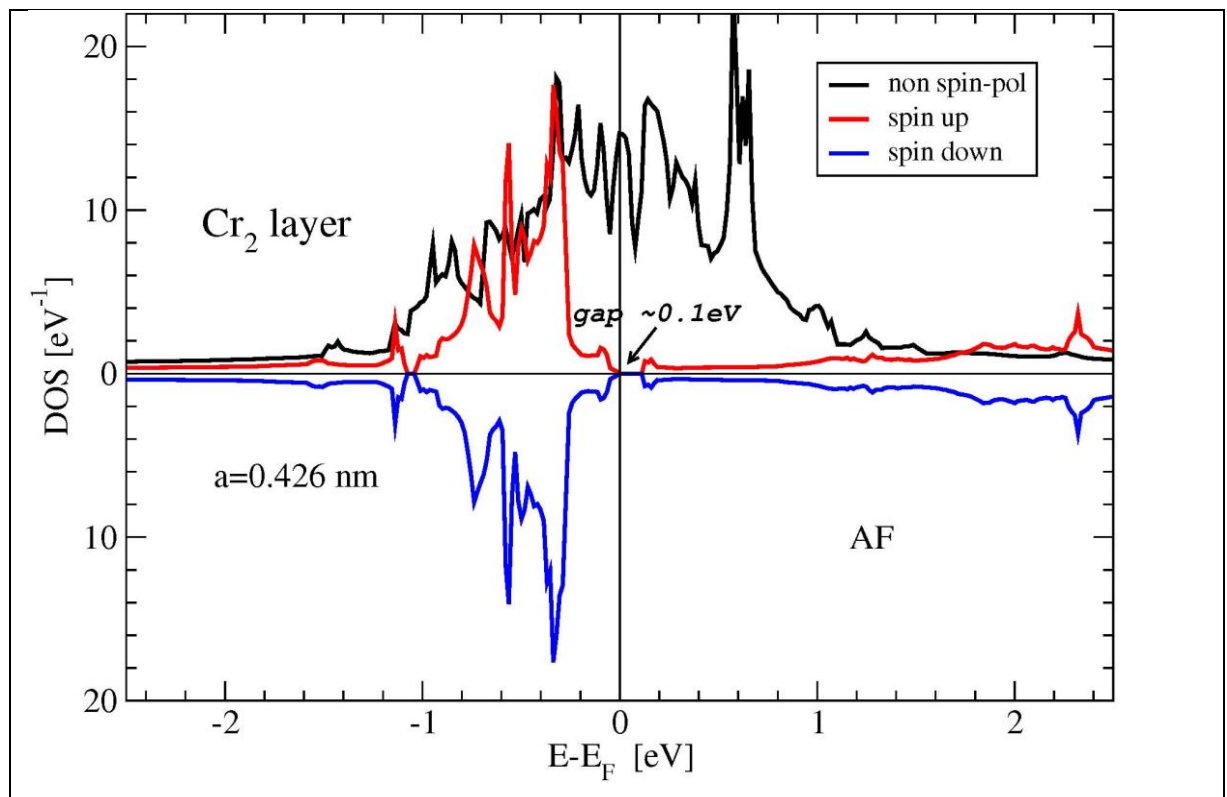

Figure S2a: Antiferromagnetic total spin up/down DOS for a Cr2 monolayer of the *P*4/*nmm* structure with lattice parameter *a* of the DFT calculation for Cr$_2$Zn$_2$Sb$_2$ calculation. Also shown is the non spin-polarised total DOS.

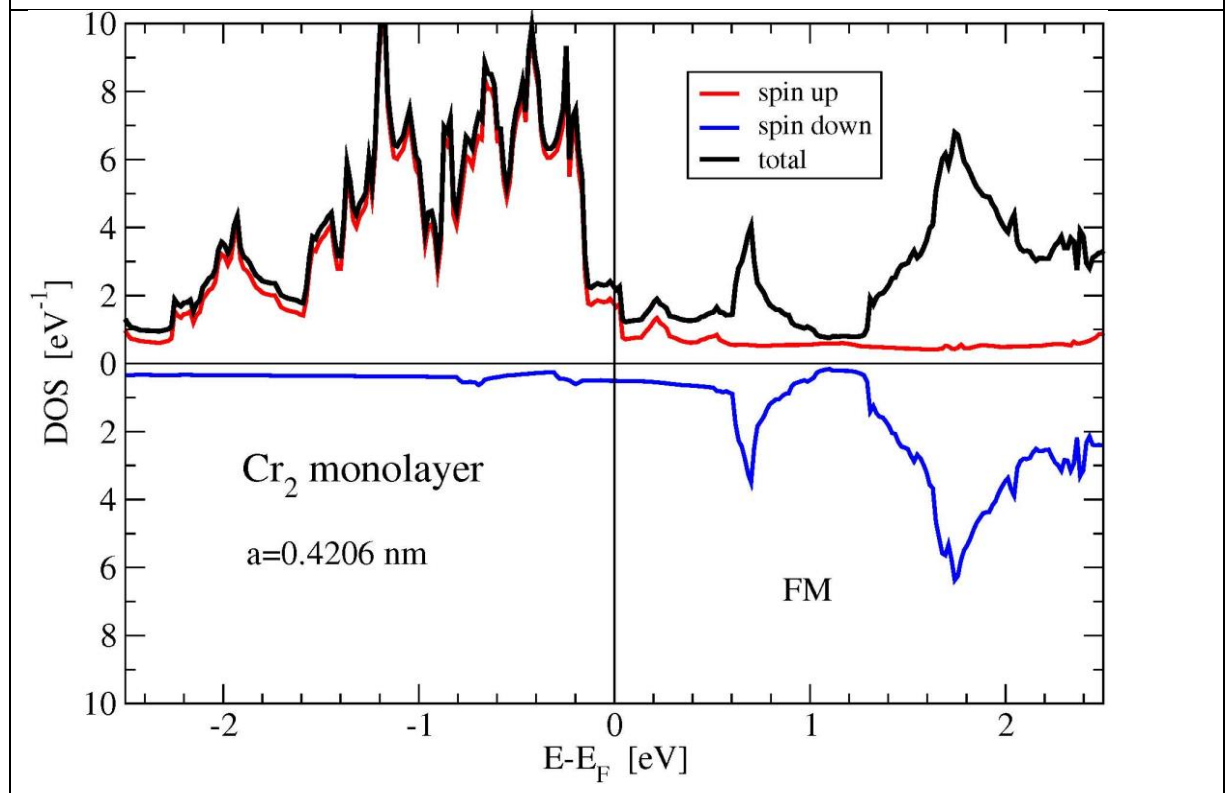

Figure S2b: Ferromagnetic local spin up/down and total spin summed DOS for a Cr$_2$ monolayer of the *P*4/*nmm* structure.

**Specific heat measurements**

At low temperatures, the specific heat follows a phenomenological temperature behavior (equ. 1)

$$C_\mathrm{p} \approx \gamma T + \beta T^3, \qquad (1)$$

which is highlighted in the $C_\mathrm{p}/T$ vs. $T^2$ plot shown in the inset of Fig. 7 by the solid line indicating a fit to the experimental data at $T^2 < 40$ K$^2$. The resulting $T$-linear coefficient $\gamma = 10.7(5)$ mJ.mol$^{-1}$K$^{-2}$ is assigned to the electronic Sommerfeld coefficient, while in the present case of Cr$_{0.86}$ZnSb, the coefficient $\beta = 0.34(2)$ mJ.mol$^{-1}$K$^{-4}$ cannot be simply assigned to lattice contributions, because antiferromagnetic spin-wave contributions may as well contribute with a $T^3$ dependence.

The electronic Sommerfeld coefficient provides a measure for the DOS at $E_\mathrm{F}$ of a metallic system and is given according to the Sommerfeld model [50] by $\gamma_S = \frac{\pi^2}{3} k_B^2 N(E_F)$, involving the electronic DOS at $E_\mathrm{F}$, $N(E_\mathrm{F})$ and Boltzmann constant $k_\mathrm{B}$. Given $N(E_\mathrm{F})$ in units of eV$^{-1}$/f.u., one obtains $\gamma_S = 2.3571 \times N(E_F)$, resulting in values (in mJ.mol$^{-1}$K$^{-2}$) of 5.15 (C-type AF CrZnSb), 4.24 (G-type AF CrZnSb), 5.82 (*vac*1-2 G-type AF Cr$_{0.875}$ZnSb), and 3.41 (*vac*1-2 G-type AF Cr$_{0.875}$ZnSb $N(E_\mathrm{F,dop})$). The energy $E_\mathrm{F,dop}$ refers to a shifted Fermi energy, obtained by doping the system with 1.46 electrons (see Section Transport, red line in Fig. 4b). All these calculated values are significantly smaller than $\gamma_\mathrm{exp} = 10.7$ mJ.mol$^{-1}$K$^{-2}$, in particular for Cr$_{0.875}$ZnSb compounds. This leads to the conclusion that an electron effective mass enhancement due to spin fluctuations may play an important role, which was discussed earlier in theory [51], as well as experiments, see e.g. Refs. [52, 53].

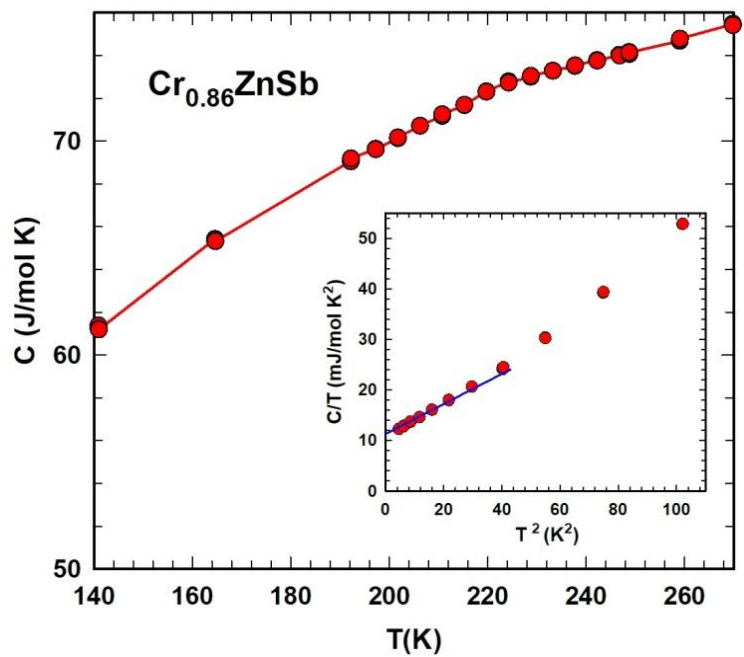

Figure S3: Specific heat, C vs. T, of Cr$_{0.86}$ZnSb in the temperature range of 140-270 K, which zooms on the antiferromagnetic transition at T$_N$ ~ 220 K. Inset: Low temperature heat capacity as C/T vs. T², where the electronic Sommerfeld coefficient is fitted as explained in the text.

It is noteworthy, that despite the slightly different nominal stoichiometries of the investigated samples, both the magnetic susceptibility and heat capacity results match almost perfectly well with corresponding data presented in the supplement of Ref. [22].

**Magnetoresistance**

The samples show negligible magnetoresistance over the whole temperature range as showcased in Fig. S3. Inspecting the comparison of the temperature-dependent resistivity at a magnetic field of 0 and 9 T, no difference in the temperature-dependent behavior can be observed.

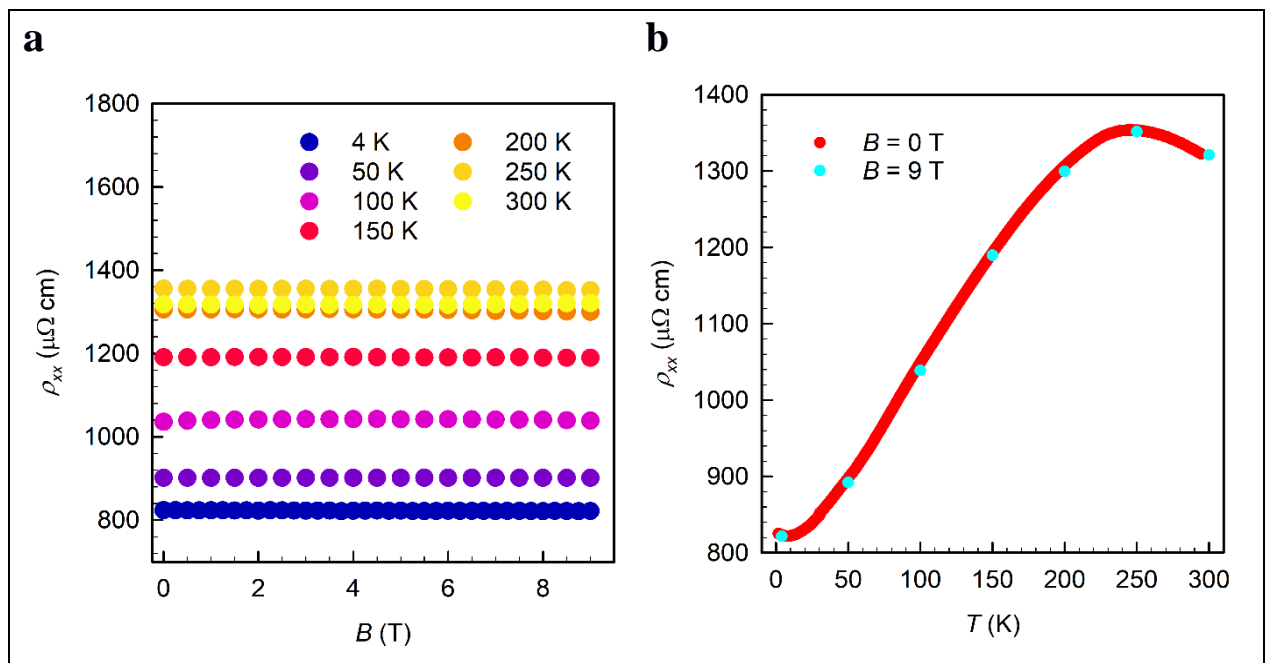

Figure S4: Panel a): Field-dependent resistivity of Cr$_{0.86}$ZnSb for different temperatures, ranging from 4-300 K. Panel b) depicts the temperature-dependent resistivity of Cr$_{0.86}$ZnSb for $B = 0$ T and $B = 9$ T in red and cyan, respectively. Evidently, the measurements reveal negligibly small magnetoresistivity.

**Zn on a Cr vacancy site**

Additional transport calculations are shown in Figure S4 and S5 for one Zn atom placed at a spin-up Cr vacancy site of the *vac*1-2 G-AF, resulting in the composition Cr$_{14}$Zn$_{17}$Sb$_{16}$. Seebeck coefficients and electronic resistivities were calculated directly from the DFT data by using BoltzTrap [6].

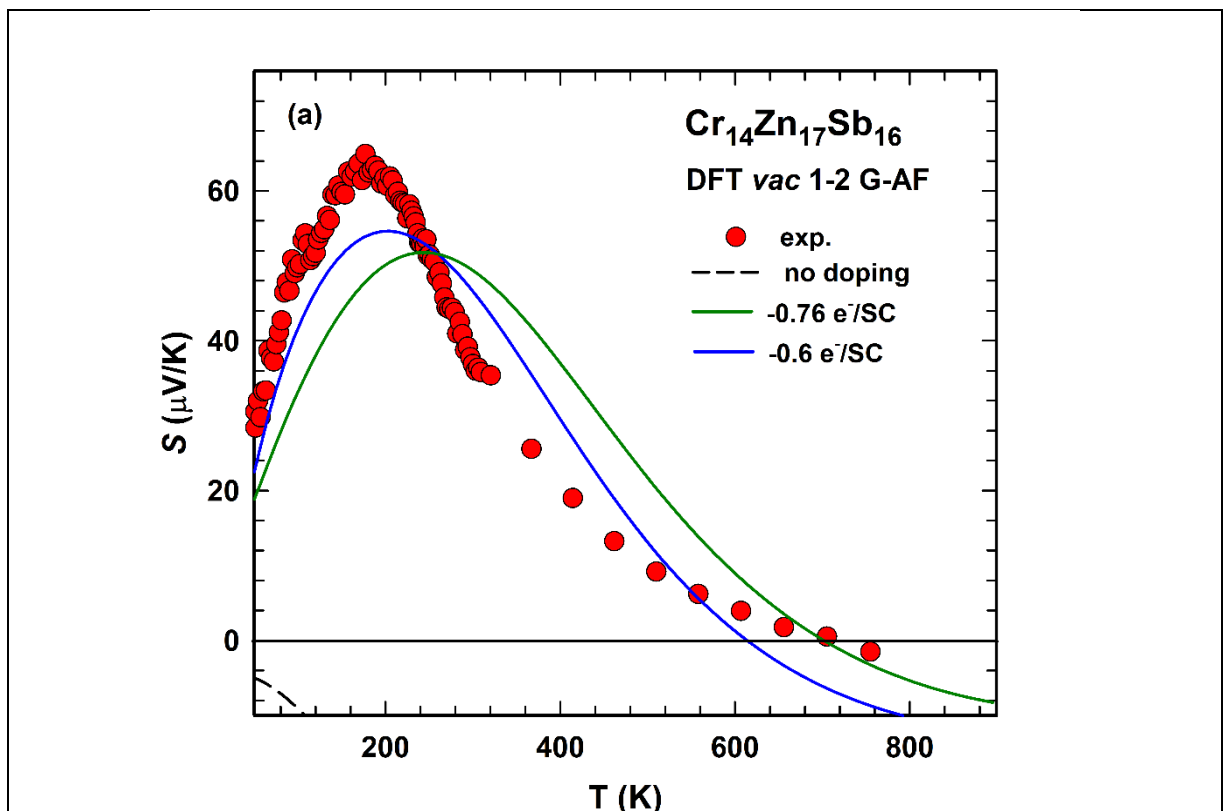

Figure S5: Seebeck coefficients for Cr$_{16}$Zn$_{17}$Sb$_{16}$ for the two positive doping levels, removing 0.6 and 0.76 electrons supercell.

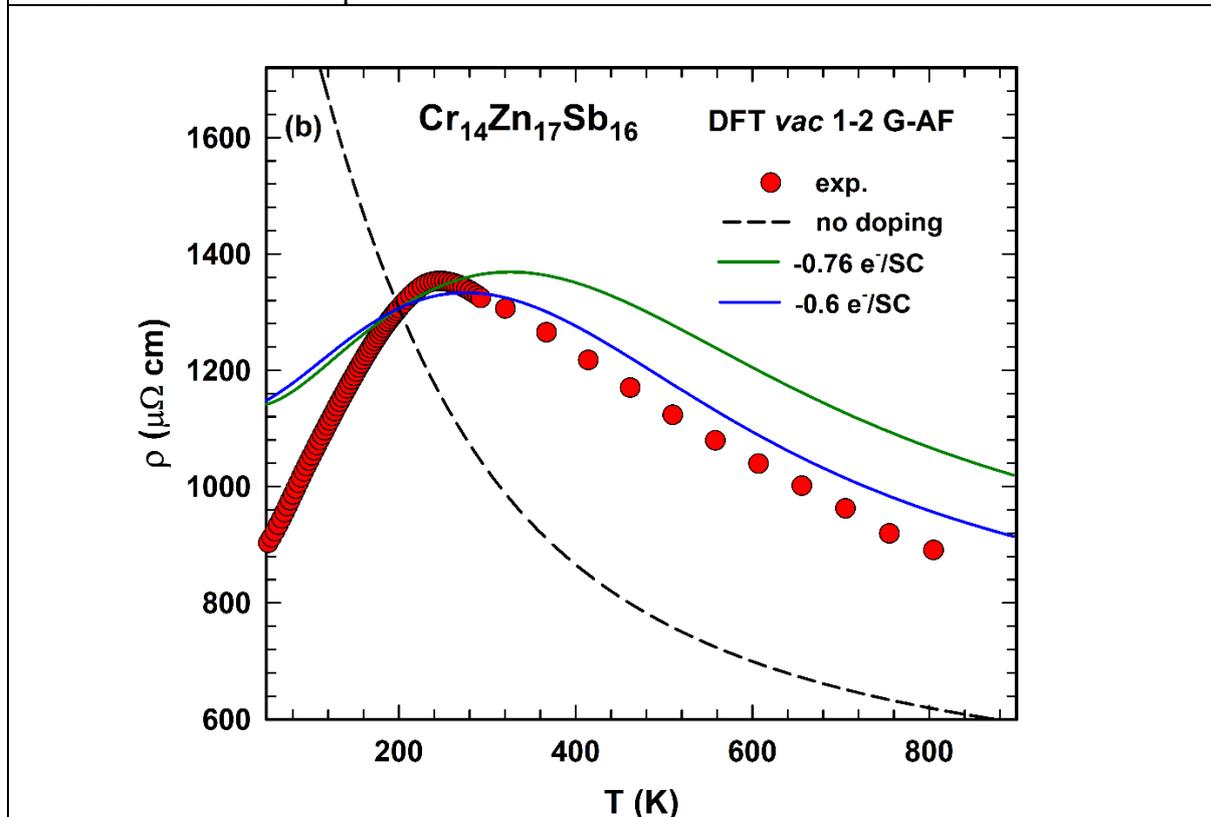

Figure S6: Electronic resistivities for Cr$_{16}$Zn$_{17}$Sb$_{16}$ for the two positive doping levels, removing 0.6 and 0.76 electrons per supercell.